\documentclass[12pt]{article}
\usepackage{amsfonts}
\usepackage{graphicx}
\usepackage{amsmath}
\usepackage{float}
\textwidth=6.5in \hoffset=-.75in \voffset=-1in \numberwithin{equation}{section}
\numberwithin{figure}{section}

\newcommand {\nn}{\nonumber}
\newcommand {\be}{\begin{equation}}
\newcommand {\ee}{\end{equation}}
\newcommand {\bea}{\begin{eqnarray}}
\newcommand {\eea}{\end{eqnarray}}
\usepackage{amsfonts}
\usepackage{amsmath}
\usepackage{amssymb}
\usepackage{graphicx}
\usepackage{float}
\usepackage{color}

\begin{document}

\begin{titlepage}
\vspace{1cm}
\begin{center}
{\Large \bf {M-Branes on Minimal Surfaces}}\\
\end{center}
\vspace{2cm}
\begin{center}
{A. M. Ghezelbash\footnote{amg142@mail.usask.ca}}
\\
Department of Physics and Engineering Physics, \\ University of Saskatchewan, \\
Saskatoon, Saskatchewan S7N 5E2, Canada\\
\vspace{1cm}
\vspace{2cm}

\end{center}

\begin{abstract}

We construct some new brane solutions in M-theory, based on the minimal surfaces. In particular, we consider the anti-self-dual Nutku geometry, and embed it in the membranes and five-branes of the eleven-dimensional supergravity. We explicitly show that the solutions preserve eight supersymmetries. Upon compactification on a circle, we find fully localized intersecting brane systems. We also discuss the T-dual and the decoupling limits of the solutions.

\end{abstract}
\end{titlepage}\onecolumn 
\bigskip 

\section{Introduction}
There are a lot of interests to find the classical soliton states of the M-theory, using the eleven dimensional supergravity, as the effective low energy limit of the M-theory \cite{gr1,gr2,gr3}.  These solutions, after compactification to ten dimensions, generate the 
supersymmetric brane systems, in which one brane is fully or partially localized on the world-volume of the other brane. Several supersymmetric brane systems, involving two or three orthogonally intersecting
2-branes and 5-branes in supergravity, were found in  
\cite{Tsey,oth}.

Moreover, some other supergravity solutions for localized D2 and D6, D2 and D4,
NS5 and D6 and NS5 and D5 intersecting brane systems, were found in \cite%
{Ch}-\cite{GH2resolvedconifolds}. The solutions were constructed by placing membranes or 5-branes in
some self-dual/anti-self-dual instanton geometries. The instanton solutions are obtained by reduction of the complex elliptic Monge-Amp\`ere equation on a complex manifold of dimension 2, which contains one real variable \cite{six}. The instanton solutions are the important parts of  constructing the exact solutions in the higher-dimensional modified theories of gravity \cite{seven}-\cite{seven3}, and in supergravity theories \cite{eight,nine}. The solutions also have been used in studying the quantum properties of the black holes \cite{ten}.
 
The self-dual geometries have been used in \cite{Ch} to construct the fully localized D2 branes intersecting the D6 branes, in type IIA string theory. The D-brane solutions were obtained by compactifying the M-brane solutions, over a circle of the transverse self-dual geometries. The metric  function for the M-brane is made of convoluted integrals of two special functions.  The localized intersecting brane solutions are supersymmetric, and exist near, as well as far from the core of the D6 branes. 

The convoluted structure for the metric function also appears in the five and higher dimensional Einstein-Maxwell theory with and without the cosmological constant \cite{twelve}. 
 
The gravitational instantons are related to the minimal surfaces in Euclidean space \cite{thirteen}, where any two-dimensional minimal surface is a solution to the real elliptic Monge-Amp\`ere equation. 
The minimal surfaces lead to the K$\ddot{\text a}$hler metrics for some gravitational instantons.

In this article, inspired by the gravitational instantons from the minimal surfaces, we construct the fully
localized brane solutions of D2 (and NS5) intersecting D6 branes. We find the type IIA brane solutions by compactification of the membrane and 5-branes of M-theory, along a Killing direction of the uplifted minimal surfaces.

The paper is organized, as follows. 

In section \ref{sec:MS}, we
discuss briefly 
the minimal surfaces and especially the Nutku geometry. We explore some properties of the Nutku space.

In section \ref{sec:M2}, we present the M2 brane solutions for the embedded Nutku space in transverse geometry, and explicitly find the number of preserved supersymmetry.  We explicitly present the type IIA brane system of D2 and D6, upon compactification over a compact direction of the Nutku geometry.
 
In sections \ref{sec:second}, we present a second class of solutions for an M2-brane, where the transverse space includes the Nutku space. We discuss about the number of preserved supersymmetry, as well as the type IIA brane system of D2 and D6, upon compactification over a compact direction of the Nutku geometry.
 
In section \ref{sec:M5}, we construct an M5-brane solution, where the transverse space includes the Nutku geometry.  We find a brane system of NS5 and D6 branes, upon compactification over a compact direction of the Nutku geometry. We explicitly find the number of preserved supersymmetry for the solution. We also consider the T-duality transformation along a parallel direction of the NS5 brane, and explicitly construct a type IIB brane system of partially intersecting NS5 and D5 branes.

In section \ref{sec:DL}, we consider the decoupling limit of the membrane and 5-brane 
solutions. We find that in the limit of vanishing string coupling,
the theory on the world-volume of the NS5-branes is a new little string
theory. Moreover, we apply T-duality transformations on type IIA solutions
and find type IIB intersecting brane solutions and discuss the
decoupling limit of the solutions. We conclude the article by three appendices, concluding
remarks and the future research directions.
\section{Minimal surfaces and the Nutku geometry }
\label{sec:MS}
The study of minimal surfaces in physics is related to almost everyday experience with a bounded layer of soap, attaching to a boundary curve. The exact definition of a minimal surface states that a surface in three-dimensional Euclidean space is minimal if and only if the surface is a critical point of the area functional. The area functional is for all the possible compact layers describe the minimal area within an existing rigid boundary, such as the surface extended by a soap film bounded on a wire frame. Other examples for minimal surfaces include the plane, the helicoid and the catenoid. The helicoid and the catenoid are locally isometric, and are harmonic conjugates of each other \cite{seven}.
The metrics for helicoid and catenoid are given by
\be
ds_{Nutku}^2=\frac{{d{{r}}}^{2}+ \left( \epsilon{N}^{2}+{r}^{2} \right) {d{{\theta}}}^{2}+
\left( 1+{\epsilon\frac {{N}^{2} \sin ^2 \theta 
}{{r}^{2}}} \right) {d{{x}}}^{2}-{\epsilon\frac {{N}^{2}\sin \left( 2\,
\theta \right) d{{x}}d{{z}}}{{r}^{2}}}+ \left( 1+{\epsilon\frac {{N}^{2}
\cos ^2 \theta}{{r}^{2}}} \right) {d
{{z}}}^{2}
}{ \sqrt{1+{\epsilon\frac {{N}^{2}}{{r}^{2}}}}}\label{Nutku},
\ee
where for the helicoid, $\epsilon=1$, and for the catenoid, $\epsilon=-1$, and we call $N$ as the Nutku parameter. The metric \eqref{Nutku} is asymptotically Euclidean, where the radial coordinate belongs to the interval $[0,+\infty[$ for the helicoid and to the interval $[N,+\infty[$ for the catenoid. The angular coordinate $\theta$ belongs to the interval $0\leq\theta\leq2\pi$ for both the helicoid and catenoid. 
The Ricci scalar and the Ricci tensor of the metric \eqref{Nutku} are identically zero, while the Kretschmann invariant is given by 
\begin{equation}
{\cal K}=\frac{72N^4}{r^4(r^2+\epsilon N^2)^2}+\frac{24N^8}{r^6(r^2+\epsilon N^2)^3},
\end{equation}
respectively. We notice that the helicoid ($\epsilon=1$) has a singularity at $r=0$, while the catenoid ($\epsilon=-1$), has another singularity at $r=N$. 


\section{Embedding Nutku geometry in M-theory}
 \label{sec:M2}
\begin{figure}[H]
\centering
\includegraphics[width=0.4\textwidth]{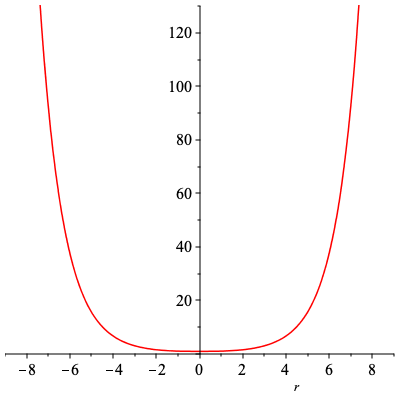}\includegraphics[width=0.4\textwidth]{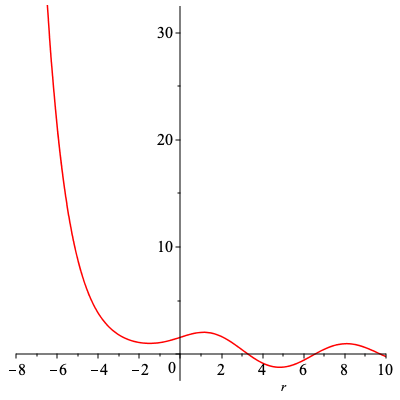}
\caption{The radial function $R(r)$, where we set $r_2=0$ (left) and $r_1=0$ (right), with $N=2,\,c=1$ and $\epsilon=1$.}
\label{R1}
\end{figure}

\begin{figure}[H]
\centering
\includegraphics[width=0.4\textwidth]{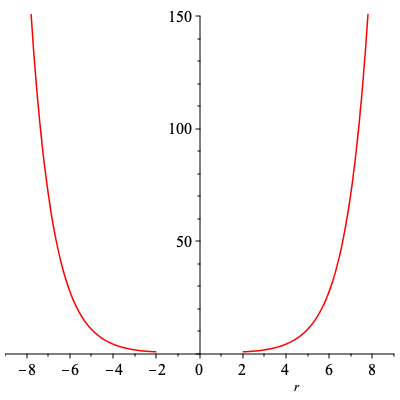}\includegraphics[width=0.4\textwidth]{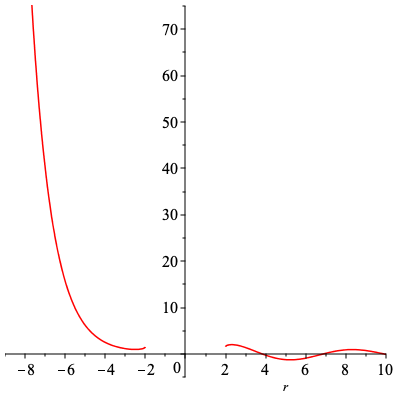}
\caption{The radial function $R(r)$, where we set $r_2=0$ (left) and $r_1=0$ (right), with $N=2,\,c=1$ and $\epsilon=-1$.}
\label{Rm1}
\end{figure}

In this section, we consider the Nutku geometry (\ref{Nutku}), as a part of transverse geometry for an M2-brane. We recall that the bosonic sector of the eleven dimensional supergravity consists of gravity and a 4-form field strength  \cite{DuffKK}. We have the following field equations
\begin{align}
G_{MN} &  =\frac{1}{3}\left[  F_{MPQR}F_{N}
^{\phantom{N}PQR}-\frac{1}{8}g_{MN}F^2\right] \label{GminGG}\\
\nabla_{M}F^{MNPQ}  &  =-\frac{1}{576}\varepsilon^{M_{1}\ldots M_{8}
NPQ}F_{M_{1}M_2M_3M_4}F_{M_{5}M_6M_7M_{8}} \label{dF}
\end{align}
where $G_{MN}=R_{MN}-\frac{1}{2}g_{MN}R $ and $F^2=F_{PQRS}F^{PQRS}$, and capital indices show the directions in the eleven-dimensional world space. We seek membrane solutions to the field equations (\ref{GminGG}) and (\ref{dF}), where the M2-brane is given by the line element
\begin{equation}
ds_{11}^{2}=H(y,r)^{-2/3}\left(  -dt^{2}+dx_{1}^{2}+dx_{2}^{2}\right)
+H(y,r)^{1/3}\left(  d{s}_{4}^{2}(y)+ds_{Nutku}^{2}
\right).  \label{ds11genM2}
\end{equation}
In (\ref{ds11genM2}), $d{s}_{4}^{2}$ stands for the four-dimensional Euclidean metric with the radial coordinate $y$, and $ds_{Nutku}^{2}$ is given by (\ref{Nutku}) with the radial coordinate $r$. We also consider the non-zero  components of the four-form field strength as
\begin{align}
F_{tx_{1}x_{2}y}  &  =-\frac{1}{2H^{2}}\frac{\partial H}{\partial y} ,
\label{Fy}\\
F_{tx_{1}x_{2}r}  &  =-\frac{1}{2H^{2}}\frac{\partial H}{\partial r}.
\label{Fr} 
\end{align}
We find all the field equations (\ref{GminGG}) and (\ref{dF}) are satisfied, if the metric function $H(y,r)$ satisfies the partial differential equation
\begin{eqnarray}
&& r\,y \, \sqrt {{ \epsilon{{N}^{2}+{r}^{2}
}}} \left( {\frac {\partial ^{2}}{\partial {y}^{2}}}H
 \left( y,r \right)  \right) +3\,r\,\sqrt {{\epsilon {{N}^{2}+{r}^{2}}}} \left( {\frac {\partial }{\partial y}}H \left( y,r \right) \right)
+y\, (\epsilon N^2+r^2)\,
 \left( {\frac {\partial ^{2}}{\partial {r}^{2}}}H \left( y,r \right) 
 \right)\nn\\
 &+&r\, y\, \left( {\frac {\partial }{
\partial r}}H \left( y,r \right)  \right)  =0.\label{PDE}
\end{eqnarray}
We solve the partial differential equation (\ref{PDE}), using the separation of variables
\begin{equation}
H(y,r)=1+Q_{M2}R(r)Y(y), \label{Hyrsep} 
\end{equation}
where $Q_{M2}$ is the charge of the M2-brane. We find two differential equations for $R(r)$ and $Y(y)$ as
\be
\frac{d^2 R(r)}{dr^2}+\frac{r}{r^2+\epsilon N^2}\frac{dR(r)}{dr}-\frac{rc^2}{\sqrt{r^2+\epsilon N^2}}R(r)=0,\label{Req}
\ee
and 
\be
y\frac{d^2}{dy^2}Y(y)+3\frac{d}{dy}Y(y)+c^2yY(y)=0.\label{Yeq}
\ee
We find two independent solutions to equation (\ref{Req}), are given by
\begin{eqnarray}
R(r)&=&r_1\, {\cal H}_D(0,0,-\epsilon N^2c^2,0,\frac{\sqrt{\epsilon N^2+r^2}}{r})\nn\\
&+&r_2\, {\cal H}_D (0,-\frac{c^2(N^4-1)}{4},\frac{c^2(N^4+1)}{2},-\frac{c^2(N^4-1)}{4},\frac{N^4+4r\sqrt{r^2+\epsilon N^2}-1}{N^4-2\epsilon N^2-4r^2+1}),\nn\\
&&\label{Rsol}
\end{eqnarray}
where ${\cal H}_D$ is the Heun-D functions, and $r_1$ and $r_2$ are constants of integration. In figure \ref{R1}, we plot the radial solutions (\ref{Rsol}), where we set $N=2,\, c=1$ and $\epsilon=1$.  Moreover, in figure \ref{Rm1}, we plot the radial solutions (\ref{Rsol}), where we set $N=2,\, c=1$ and $\epsilon=-1$.  As we notice from figure \ref{Rm1}, the catenoid radial solutions are not continuous, and not defined over the range of coordinate $r$, due to the presence of $\sqrt{r^2-N^2}$, in equation (\ref{Rsol}). Hence it what follows, we consider only the helicoid solutions with $\epsilon=1$.

Moreover the solutions to equation (\ref{Yeq}) are given by
\be
Y(y)=y_1\frac{J_1(cy)}{y}+y_2\frac{Y_1(cy)}{y},\label{Ysol}
\ee
in terms of the Bessel functions, where $y_1$ and $y_2$ are constants of integration. Furnished with the solutions to equation (\ref{PDE}) for any separation constant $c$, we find the most general solution to (\ref{PDE}), by superimposing all solutions with different $c$, as
\begin{eqnarray}
H(y,r)&=&1+Q_{M2}\int _0 ^\infty dc \{ {\cal H}_D(0,0,- N^2c^2,0,\frac{\sqrt{ N^2+r^2}}{r})\times(f_1(c)\frac{J_1(cy)}{y}+f_2(c)\frac{Y_1(cy)}{y})\nonumber\\
&+&{\cal H}_D (0,-\frac{c^2(N^4-1)}{4},\frac{c^2(N^4+1)}{2},-\frac{c^2(N^4-1)}{4},\frac{N^4+4r\sqrt{r^2+ N^2}-1}{N^4-2 N^2-4r^2+1})\nonumber\\
&\times&(f_3(c)\frac{J_1(cy)}{y}+f_4(c)\frac{Y_1(cy)}{y})\},\label{GEN1}
\end{eqnarray}
where $f_i(c)$ with $i=1,\cdots ,4$ are four arbitrary functions of the separation constant. We find the functions $f_i(c)$ by considering the near horizon limit. Though it appears $Y_1$ in (\ref{GEN1}) is divergent at the lower limit of integral, however, we will find that the function $f_4(c)=0$, and so there is no contribution from $Y_1$ in (\ref{GEN1}). In the limit of $N \rightarrow 0$, the four-dimensional metric (\ref{Nutku}) reduces to $D^2 \times R^2$, where $D^2$ is a two-dimensional disk. In this limit, we find an exact solution to the field equation (\ref{GminGG}) and (\ref{dF}),  as
\begin{equation}
{\widehat{ds}}_{11}^{2}=\hat H_0(y,r,x,z)^{-2/3}\left(  -dt^{2}+dx_{1}^{2}+dx_{2}^{2}\right)
+\hat H_0(y,r,x,z)^{1/3}\left(  d{s}_{4}^{2}(y)+ds_{D^2 \times R^2}^{2}
\right),  \label{ds11flatM2}
\end{equation}
where 
\be
\hat H_0(y,r,x,z)=1+\frac{Q_{M2}}{(r^2+x^2+y^2+z^2)^{3}}.\label{H0}
\ee
The non-zero components of the four-form field strength are still given by (\ref{Fy}) and (\ref{Fr}), with replacement $H$ by $\hat H_0$. As we are interested to find the unique membrane solutions in the bulk, which covers from the outside of horizon to the far infinity, we should satisfy two boundary conditions on (or near) the horizon and infinity.  Hence, we require the metric function (\ref{GEN1}) reduces to (\ref{H0}), in the appropriate near horizon limit, where the M2-brane is located at $x=z=0$. We find an integral equation for $f_i(c)$ ($i=1,\cdots ,4$), which is given by
\begin{eqnarray}
&\lim_{N\rightarrow 0}&\int _0 ^\infty dc \{ {\cal H}_D(0,0,- N^2c^2,0,\frac{\sqrt{ N^2+r^2}}{r})\times(f_1(c)\frac{J_1(cy)}{y}+f_2(c)\frac{Y_1(cy)}{y})\nonumber\\
&+&{\cal H}_D (0,-\frac{c^2(N^4-1)}{4},\frac{c^2(N^4+1)}{2},-\frac{c^2(N^4-1)}{4},\frac{N^4+4r\sqrt{r^2+ N^2}-1}{N^4-2 N^2-4r^2+1})\nonumber\\
&\times&(f_3(c)\frac{J_1(cy)}{y}+f_4(c)\frac{Y_1(cy)}{y})\}=\frac{1}{(r^2+y^2)^{3}}.\label{INT}
\end{eqnarray}
To solve (\ref{INT}) and find $f_i(c)$, we notice the limits of the Heun-D functions can be obtained by looking at the solutions to the radial equation (\ref{Req}), where $N \rightarrow 0$. In fact, we find the limit of the first Heun-D function in (\ref{INT}), is proportional to the Bessel function $I_0(cr)$, while the limit of the second Heun-D function in (\ref{INT}), is proportional to the Bessel function $K_0(cr)$. We then find the unique solutions to the integral equation (\ref{INT}), which are given by
\be
f_1(c)=0,\,f_2(c)=0,\,f_3(c)=-\frac{1}{16} c^4,\,f_4(c)=0.\label{fs}
\ee
To summarize, the metric function for the M2-brane solution (\ref{ds11genM2}) is 
\begin{eqnarray}
H(y,r)&=&1-\frac{Q_{M2}}{16}\int _0 ^\infty dc c^4 \frac{J_1(cy)}{y}  {\cal H}_D (0,-\alpha{c^2},(1+2\alpha)c^2,-\alpha c^2,\frac{4\alpha+4r\sqrt{r^2+ N^2}}{2(1+2\alpha-2r^2-N^2)})
,\nonumber\\
&&\label{GENFFIN}
\end{eqnarray}
where $\alpha=\frac{N^4-1}{4}$. We note that in asymptotic limits, where $y \rightarrow \infty$, or $r \rightarrow \infty$, or both, the integrand is a superposition of two decaying oscillating functions. Hence the metric function $H(y,r)$ asymptotically approaches the constant value of 1.
{\textcolor{black}{Although we can't find an analytic expression for the integral in (\ref{GENFFIN}), we numerically find the behaviour of the metric function, which is shown in figure \ref{fig:HM2}. The left figure in \ref{fig:HM2} shows the logarithm of  $H(y=0,r)-1$ versus logarithm of $\frac{r}{N}$. Moreover, the right figure in \ref{fig:HM2} shows the logarithm of  $H(y,r=0)-1$ versus logarithm of $\frac{y}{N}$. We note that in the extremal asymptotic limit, where  $r \rightarrow \infty$ or  $y \rightarrow \infty$, the logarithm of $H(y=0,r)-1$ or $H(y,r=0)-1$ approaches $-\infty$, hence the metric function asymptotically approaches the constant value of 1.  On the other extremal limit, i.e. in the near core limit where $r \rightarrow 0$ and $y \rightarrow 0$, the logarithm of $H(y=0,r)-1$ or $H(y,r=0)-1$ approaches the constant value of 0. Comparing the numerical solutions also shows that the metric function approaches asymptotically to the constant value of 1 at shorter distance $r$, and longer distance in $y$-direction.}

\begin{figure}[H]
\centering
\includegraphics[width=0.4\textwidth]{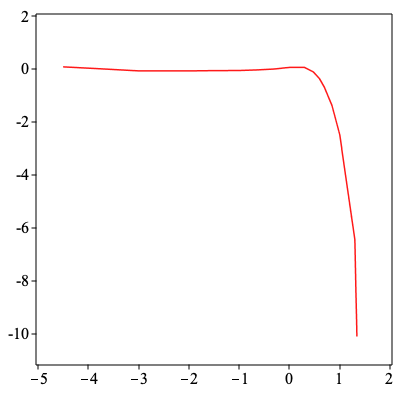}\includegraphics[width=0.4\textwidth]{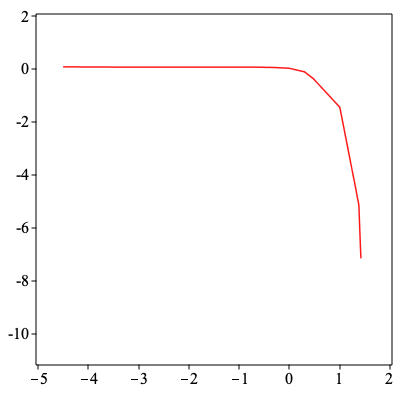}
\caption{The numerical solutions for the logarithm of  $H(y=0,r)-1$ versus logrithm of $\frac{r}{N}$ (left) and the logarithm of  $H(y,r=0)-1$ versus logrithm of $\frac{y}{N}$, where we set $N=2$.}
\label{fig:HM2}
\end{figure}

Considering either the $x$ or $z$ directions to be periodic with period $2\pi$, we find the different fields in ten dimensions, by using either the Killing vector $\partial / \partial x$ or  $\partial / \partial z$. For the Killing vector $\partial / \partial x$, we find
\begin{eqnarray}
ds_{10}^2&=&\frac{H^{-1/2}}{(1+\frac{N^2}{r^2})^{1/4}}\sqrt{{1+\frac{N^2\sin ^2\theta}{r^2}}}(-dt^2+dx_1^2+dx_2^2)+\frac{H^{1/2}}{(1+\frac{N^2}{r^2})^{1/4}}\sqrt{{1+\frac{N^2\sin ^2\theta}{r^2}}}ds_4^2(y)\nonumber\\
&+&\frac{H^{1/2}}{(1+\frac{N^2}{r^2})^{3/4}}\sqrt{{1+\frac{N^2\sin ^2\theta}{r^2}}}({{d{{r}}}^{2}+ \left({N}^{2}+{r}^{2} \right) {d{{\theta}}}^{2}+
 \left( 1+{\frac {{N}^{2}
\cos ^2 \theta}{{r}^{2}}} \right) {d
{{z}}}^{2}
}).\label{ds10}
\end{eqnarray}
The metric (\ref{ds10}) shows a D2 brane, which is fully localized along the world-volume of another D6 brane. 
Moreover the only non-zero NSNS field is the dilaton field, which is given by
\be
\Phi=\frac{3}{4}\ln \{\frac{(1+\frac{N^2\sin^2\theta}{r^2})H^{1/3}}{\sqrt{1+\frac{N^2}{r^2}}}\}.\label{dilIIA}
\ee
The RR fields are 
\be
C_z=-\frac{N^2\sin \theta \cos\theta}{r^2+{N^2\sin^2\theta}},\label{CIIA}
\ee
and
\be
A_{tx_1x_2}=\frac{1}{H}.\label{A3}
\ee
We note that in equations (\ref{ds10})-(\ref{A3}), $H$ is given  by (\ref{GENFFIN}). We have explicitly verified that the metric and different fields (\ref{ds10})-(\ref{A3}) satisfy exactly all the ten-dimensional type IIA supergravity field equations.  

We find the number of preserved supersymmetry by finding the non-trivial solutions to the Killing spinor equation, for the anticommuting parameter $\epsilon$, of the supersymmetry transformations.  We note that the Killing spinor equation, is equivalent to the variation of the gravitino field in the eleven-dimensional M-theory.  The field equations (\ref{GminGG}) and (\ref{dF}), are for the bosonic fields of the eleven-dimensional M-theory, and so the supersymmetric variation of the gravitino must vanish. The vanishing of the supersymmetric variation of the gravitino, yields the Killing spinor equation, which is given by
\be
(\partial_M + \frac{1}{4}\omega_{M\mu\nu}\Gamma ^{\mu\nu}+\frac{1}{144}\Gamma _{M}^{NPQR}F_{NPQR}-\frac{1}{18}\Gamma^{PQR}F_{MPQR})\epsilon=0,\label{K}
\ee
where $\mu,\,\nu$ are the eleven dimensional tangent space indices. 
\label{sec:aneqb}
In (\ref{K}), we have the following 
\begin{eqnarray}
\omega ^\mu_{\nu \rho}&=&\frac{1}{2}(h^\mu _{\nu\rho}+h^\nu _{\rho\mu}-h^\rho _{\mu\nu}),\label{omeg}\\
\omega_{\mu\nu M}&=&\omega ^{\rho}_{\nu\lambda}\eta_{\mu\rho}e^{\lambda}_M,\label{omeg2}
\end{eqnarray}
where the elfbein $e^\mu=e^\mu_M\,dx^M$ satisfy
\begin{eqnarray}
de^\mu&=&h^{\mu}_{\nu\rho}e^\nu \wedge e^\rho,\label{de}\\
g_{MN}&=&\eta_{\mu\nu}e^\mu _M e^\nu_N.\label{gMN}
\end{eqnarray}
{\textcolor{black}{In appendix \ref{app}, we present the explicit dependence of $e_{\mu M}$, $\omega_{\mu\nu M}$ and $\omega^\mu_{\nu \rho}$ on the metric function $H$ and its derivatives.}
The different types of $\Gamma$'s in the Killing spinor equation (\ref{K}) are
\begin{eqnarray}
\Gamma^{\mu\nu}&=&\Gamma^{[\mu}\Gamma^{\nu]},\label{gam2}\\
\Gamma^{M_1\cdots M_n}&=&\Gamma^{[M_1}\cdots \Gamma^{M_n]}.\label{gamn}
\end{eqnarray}
We note that $\Gamma^\mu$ satisfies the Clifford algebra
\be
\{\Gamma ^\mu,\Gamma^\nu\}=-2\eta^{\mu\nu}.\label{GAMmu}
\ee
We use the representation for the $\Gamma ^\mu$, as
\begin{eqnarray}
\Gamma _\aleph&=&\gamma_\aleph \otimes \mathbb{I}_8,\label{anti1}\\
\Gamma _{\beth+4}&=&\gamma_5 \otimes \widehat \Gamma_\beth,\label{GAM}
\end{eqnarray}
where $\aleph=0,\cdots,3$ denotes the indices of the tangent space group $SO(1,3)$, and $\beth=0,1,\cdots,6$ denotes the indices of the  tangent space group $SO(7)$, respectively. Both sets of $\Gamma _{\beth+4}$ and $\widehat \Gamma _{\beth}$ satisfy the anticommutation algebra
\begin{eqnarray}
\{\Gamma _{\beth+4},\Gamma _{\beth'+4}\}&=&-2\delta_{\beth \beth'},\label{anti3}\\
\{\widehat\Gamma _{\beth},\widehat\Gamma _{\beth'}\}&=&-2\delta_{\beth \beth'}.
\end{eqnarray}
In (\ref{GAM}), $\gamma_\aleph$ are the Dirac matrices,  $\gamma_5=i\gamma_0\gamma_1\gamma_2\gamma_3$ and $\mathbb{I}$ is the identity group element. Moreover, we consider the following representation for the $\widehat \Gamma _{\beth}$,
\begin{eqnarray}
\widehat \Gamma _{0}&=&i\gamma_0 \otimes \mathbb{I}_2,\\
\widehat \Gamma _{i}&=&\gamma_i \otimes \mathbb{I}_2,\\
\widehat \Gamma _{i+3}&=&i\gamma_5\otimes \sigma_i,
\end{eqnarray}
where $\sigma_i,\,i=1,2,3$ are the Pauli matrices. We also mention another useful representation for the Clifford algebra (\ref{GAMmu}) \cite{20},
\begin {eqnarray}
\Gamma_0&=&-\Gamma_{123456789\sharp},\label{oct1}\\
\Gamma_\sharp&=&\begin{bmatrix}
0 &   \mathbb{I}_{16}\\
	\mathbb{I}_{16} & 0 \\
\end{bmatrix},\\
\Gamma_9&=&\begin{bmatrix}
 \mathbb{I}_{16} &0\\
0&-\mathbb{I}_{16} \\
\end{bmatrix},\\
\Gamma_\daleth&=&\begin{bmatrix}
 0&-\widetilde \Gamma_\daleth\\
\widetilde \Gamma_\daleth&0\\
\end{bmatrix},\label{oct4}
\end{eqnarray}
where $\widetilde \Gamma_\daleth,\,\daleth=1,\cdots ,8$ are the sixteen dimensional matrix representation of the Clifford algebra in eight dimensions. They are given by
\be
\widetilde\Gamma_i=\begin{bmatrix}
 0&{\cal{O}}_i\\
{\cal{O}}_i&0\\
\end{bmatrix},\label{GAMTILi}
\ee
where $i=1,\cdots,7$ and 
\be
\widetilde\Gamma_8=\begin{bmatrix}
 0&-\mathbb{I}_8\\
\mathbb{I}_8&0\\
\end{bmatrix}.
\ee
In (\ref{GAMTILi}), ${\cal{O}}_i$ represent the left multiplication operators by the imaginary octonions on the octonions. To construct them, we note that imaginary unit octonions satisfy
\be
o_i \cdot o_j = -\delta_{ij}+f_{ijk}o_k,\label{oc}
\ee
where the structure constants $f_{ijk}$ make a skew symmetric tensor. The only non zero components of $f_{ijk}$ are
$
f_{124}=f_{137}=f_{156}=f_{235}=f_{267}=f_{346}=f_{457}=1
$. If we denote a vector in eight dimensional real space by $V=(V_0,V_i)$, we have a corresponding octonion $\hat V=V_0+V_io_i$. The left multiplication of octonion $o_i$ on $\hat V$ is $o_i(\hat V)=V_0o_i-V_i+f_{ijk}V_jo_k$. We then find the eight dimensional representation of the operators ${\cal{O}}_i$ as
$o_i(\hat V)=({\cal {O}}_i)_{\hat i\hat j}o_{\hat i}V_{\hat j}$, where ${\hat i},{\hat j}=0,1,\cdots ,7$. Using equation (\ref{K}) with $M=t,x_1$ and $x_2$, we find 
\be
\Gamma ^{\hat t \hat x_1 \hat x_2}\epsilon=-\epsilon,\label{first}
\ee
where we denote the tangent space indices, with an overhead hat. 
{\textcolor{black}{In appendix \ref{app}, we present explicitly all the different terms in equation (\ref{K}) with $M=t,x_1$, and $x_2$, which leads to equation (\ref{first}).}
Equation (\ref{first}) reduces the number of independent components of the spinor $\epsilon$ by half. Using equation (\ref{K}) with $M=y$ and $r$, yields two equations which are trivially satisfied. 
{\textcolor{black}{In appendix \ref{app}, we present explicitly all the different terms in equation (\ref{K}) with $M=y$ and $r$, which leads to trivial equations.}
The equation (\ref{K}) with $M=\alpha_1,\alpha_2$ and $\alpha_3$ leads to the following equations
\begin{eqnarray}
\partial _{\alpha_1}\epsilon&=& -\frac{\Gamma^{\hat \alpha_1 \hat y}}{2}\epsilon,\label{K1}\\
\partial _{\alpha_2}\epsilon&=&\{ -\sin(\alpha_1)\frac{\Gamma^{\hat \alpha_2 \hat y}}{2}+\cos(\alpha_1)\frac{\Gamma^{\hat \alpha_1 \hat \alpha_2}}{2}\}\epsilon,\label{K2}\\
\partial _{\alpha_3}\epsilon&=&\{ \sin(\alpha_1)\sin(\alpha_2)\frac{\Gamma^{\hat y\hat \alpha_3}}{2}+\cos(\alpha_1)\sin(\alpha_2)\frac{\Gamma^{\hat \alpha_1 \hat \alpha_3}}{2}+\cos(\alpha_2)\frac{\Gamma^{\hat \alpha_2 \hat \alpha_3}}{2}\}\epsilon.\label{K3}
\end{eqnarray}
{\textcolor{black}{In appendix \ref{app}, we present explicitly all the different terms in equation (\ref{K}) with $M=\alpha_1,\alpha_2$ and $\alpha_3$, which leads to equations (\ref{K1})-(\ref{K3}).}
We find that the solution to the equations (\ref{K1}),(\ref{K2}) and (\ref{K3}), is 
\be
\epsilon=e^{-\alpha_1\frac{\Gamma^{\hat \alpha_1 \hat y}}{2}}e^{\alpha_2\frac{\Gamma^{\hat \alpha_1 \hat \alpha_2}}{2}}e^{\alpha_3\frac{\Gamma^{\hat \alpha_2 \hat \alpha_3}}{2}}\epsilon'\label{eprime}
\ee
where $\epsilon'$ is independent of $\alpha_1,\alpha_2$ and $\alpha_3$. The equation (\ref{K}) with $M=z$ leads to the following equations
\begin{eqnarray}
\partial _{z}\epsilon'&=&(Z_{\hat z \hat r} \Gamma^{\hat z\hat r}+Z_{\hat \theta \hat x} \Gamma^{\hat \theta \hat x}+Z_{\hat \theta \hat z} \Gamma^{\hat \theta \hat z}+Z_{\hat x \hat r} \Gamma^{\hat x \hat r})\epsilon',\label{K5}
\end{eqnarray}
where the different $Z$ functions are given explicitly in the appendix \ref{app.gamma}. Multiplying equation (\ref{K5}) from left by $\Gamma^{\hat \theta\hat \psi}$ and assuming $\epsilon'$ is independent of $z$, leads to
\be
\Gamma^{\hat \theta \hat z \hat x \hat r} \epsilon'=\epsilon',\label{epp}
\ee
We notice that equation (\ref{epp}) reduces the number of independent components of the spinor $\epsilon'$ by half. Equation (\ref{epp}) is a reflection of the anti-self-duality of the Nutku geometry in four dimensions. We also find that equation (\ref{K}) with $M=\theta$ and $x$, leads to the same equation (\ref{epp}). As a result, we conclude that M2 brane solution (\ref{ds11genM2}), with the metric function (\ref{GENFFIN}) preserves eight supersymmetries. {\textcolor{black}{We note from the explicit derivation of the Killing spinor equations for the M2-brane (in appendix \ref{app}), that the final projection equations (\ref{first}) and (\ref{epp}) are independent of the explicit dependence of the metric function $H(y,r)$ on the coordinates $y$ and $r$, as given by (\ref{GENFFIN}).}}
\section{A second set of M2 brane solutions}
 \label{sec:second}
 
In this section, we find another independent M2-brane solutions $\tilde H(y,r)=1+Q_{M_2}\tilde R(r)\tilde Y(y)$, by analytic continuation of the separation constant $c$ in equations (\ref{Req}) and (\ref{Yeq}), to $ic$. The solutions for $\tilde Y(y)$, are given by
\be
\tilde Y(y)=\tilde y_1\frac{I_1(cy)}{y}+\tilde y_2\frac{K_1(cy)}{y},\label{Ytildesol}
\ee
in terms of the modified Bessel functions $I_1$ and $K_1$, where $\tilde y_1$ and $\tilde y_2$ are constants of integration. Moreover, we find the solutions for $\tilde R(r)$ are given by
\begin{eqnarray}
\tilde R(r)&=&\tilde r_1\, {\cal H}_D(0,0,\epsilon N^2c^2,0,\frac{\sqrt{\epsilon N^2+r^2}}{r})\nn\\
&+&\tilde r_2\, {\cal H}_D (0,\frac{c^2(N^4-1)}{4},-\frac{c^2(N^4+1)}{2},\frac{c^2(N^4-1)}{4},\frac{N^4+4r\sqrt{r^2+\epsilon N^2}-1}{N^4-2\epsilon N^2-4r^2+1}),\nn\\
&&\label{Rtildesol}
\end{eqnarray}
where $\tilde r_1$ and $\tilde r_2$ are constants of integration. In figure \ref{R1tilde}, we plot the radial solutions (\ref{Rtildesol}), where we set $N=2,\, c=1$ and $\epsilon=1$.  Moreover, in figure \ref{Rm1tilde}, we plot the radial solutions (\ref{Rtildesol}), where we set $N=2,\, c=1$ and $\epsilon=-1$. As we notice from figure \ref{Rm1tilde}, the catenoid radial solutions are not continuous, and not defined over the range of coordinate $r$, due to the presence of $\sqrt{r^2-N^2}$, in equation (\ref{Rtildesol}). Hence in what follows, we consider only the helicoid solutions with $\epsilon=1$, similar to section \ref{sec:M2}.

\begin{figure}[H]
\centering
\includegraphics[width=0.4\textwidth]{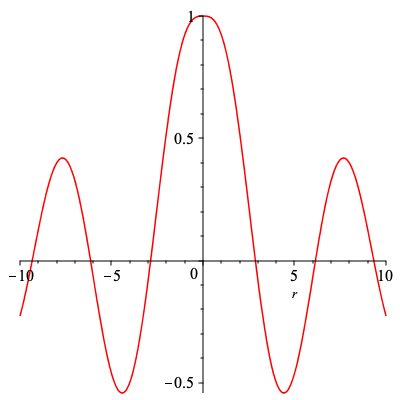}\includegraphics[width=0.4\textwidth]{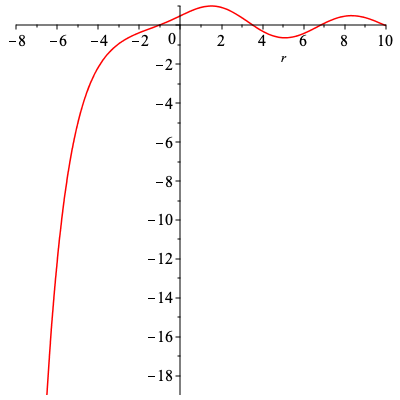}
\caption{The radial function $\tilde R(r)$, where we set $\tilde r_2=0$ (left) and $\tilde r_1=0$ (right), with $N=2,\,c=1$ and $\epsilon=1$.}
\label{R1tilde}
\end{figure}

\begin{figure}[H]
\centering
\includegraphics[width=0.4\textwidth]{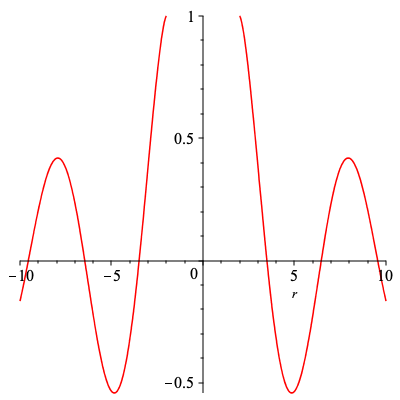}\includegraphics[width=0.4\textwidth]{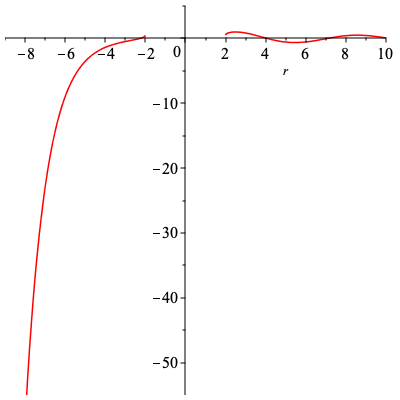}
\caption{The radial function $\tilde R(r)$, where we set $\tilde r_2=0$ (left) and $\tilde r_1=0$ (right), with $N=2,\,c=1$ and $\epsilon=-1$.}
\label{Rm1tilde}
\end{figure}

We find now that the most general solution, for the metric function $\tilde H(y,r)$ can be written as the superposition of all the solutions with different $c$, which is 
\begin{eqnarray}
\tilde H(y,r)&=&1+Q_{M2}\int _0 ^\infty dc \{ {\cal H}_D(0,0,N^2c^2,0,\frac{\sqrt{ N^2+r^2}}{r})\times(\tilde f_1(c)\frac{I_1(cy)}{y}+\tilde f_2(c)\frac{K_1(cy)}{y})\nonumber\\
&+&{\cal H}_D (0,\frac{c^2(N^4-1)}{4},-\frac{c^2(N^4+1)}{2},\frac{c^2(N^4-1)}{4},\frac{N^4+4r\sqrt{r^2+ N^2}-1}{N^4-2 N^2-4r^2+1})\nonumber\\
&\times&(\tilde f_3(c)\frac{I_1(cy)}{y}+\tilde f_4(c)\frac{K_1(cy)}{y})\},\label{GENtilde}
\end{eqnarray}
where $ \tilde f_i(c)$ with $i=1,\cdots ,4$, are four arbitrary functions of the separation constants. Though it appears $K_1$ in (\ref{GENtilde}) is divergent at the lower limit of integral, however, we will find that the function $\tilde f_4(c)=0$, and so there is no contribution from $K_1$ in (\ref{GENtilde}).

We find the functions $\tilde f_i(c)$, by considering the near horizon limit, where the four-dimensional metric (\ref{Nutku}) reduces to $D^2 \times R^2$, where $D^2$ is a two-dimensional disk. Similar to section \ref{sec:M2}, we require the metric function (\ref{GENtilde}) reduces to (\ref{H0}), in the appropriate near horizon limit, where the M2-brane is located at $x=z=0$. We find an integral equation for $\tilde f_i(c)$ ($i=1,\cdots ,4$), which is given by
\begin{eqnarray}
&\lim_{N\rightarrow 0}&\int _0 ^\infty dc \{ {\cal H}_D(0,0,N^2c^2,0,\frac{\sqrt{ N^2+r^2}}{r})\times(\tilde f_1(c)\frac{I_1(cy)}{y}+\tilde f_2(c)\frac{K_1(cy)}{y})\nonumber\\
&+&{\cal H}_D (0,\frac{c^2(N^4-1)}{4},-\frac{c^2(N^4+1)}{2},\frac{c^2(N^4-1)}{4},\frac{N^4+4r\sqrt{r^2+ N^2}-1}{N^4-2 N^2-4r^2+1})\nonumber\\
&\times&(\tilde f_3(c)\frac{I_1(cy)}{y}+\tilde f_4(c)\frac{K_1(cy)}{y})\}=\frac{1}{(r^2+y^2)^{3}}.\label{INTtilde}
\end{eqnarray}
To solve (\ref{INTtilde}) and find $\tilde f_i(c)$, we notice the limits of the Heun-D functions can be obtained by looking at the solutions (\ref{Rtildesol}), to the radial equation, where $N \rightarrow 0$. In fact, we find the limit of the first Heun-D function in (\ref{INTtilde}), is the Mathieu-C function, while the limit of the second Heun-D function in (\ref{INTtilde}), is proportional to the Bessel function $K_0(icr)$. We then find the unique solutions to the integral equation (\ref{INTtilde}), which are given by
\be
\tilde f_1(c)=0,\,\tilde f_2(c)=0,\,\tilde f_3(c)=\frac{1}{16} c^4,\,\tilde f_4(c)=0.\label{fstilde}
\ee
To summarize, the metric function for the second M2-brane solution (\ref{ds11genM2}) is 
\begin{eqnarray}
\tilde H(y,r)&=&1+\frac{Q_{M2}}{16}\int _0 ^\infty dc c^4 \frac{I_1(cy)}{y}  {\cal H}_D (0,\alpha{c^2},-(1+2\alpha)c^2,\alpha c^2,\frac{4\alpha+4r\sqrt{r^2+ N^2}}{2(1+2\alpha-2r^2-N^2)})
,\nonumber\\
&&\label{GENfinaltilde}
\end{eqnarray}
We note that in asymptotic limits, where $y \rightarrow \infty$, or $r \rightarrow \infty$, or both, the integrand is a superposition of a diverging Bessel function, and a decaying oscillating Heun-D function.  We numerically evaluate the asymptotic values of the integral in  (\ref{GENfinaltilde}). The results show that the decaying Heun-D function takes over the diverging Bessel function, and the metric function $H(y,r)$ asymptotically approaches the constant value of 1.
{\textcolor{black}{Although we can't find an analytic expression for the integral in (\ref{GENfinaltilde}), we numerically find the behaviour of the metric function, which is shown in figure \ref{fig:HMM2}. The left figure in \ref{fig:HMM2} shows the logarithm of  $\tilde H(y=0,r)-1$ versus logarithm of $\frac{r}{N}$. Moreover, the right figure in \ref{fig:HMM2} shows the logarithm of  $\tilde H(y,r=0)-1$ versus logarithm of $\frac{y}{N}$. We note that in the extremal asymptotic limit, where  $r \rightarrow \infty$ or  $y \rightarrow \infty$, the logarithm of $\tilde H(y=0,r)-1$ or $\tilde H(y,r=0)-1$ approaches $-\infty$, hence the metric function asymptotically approaches the constant value of 1.  On the other extremal limit, i.e. in the near core limit where $r \rightarrow 0$ and $y \rightarrow 0$, the logarithm of $\tilde H(y=0,r)-1$ or $\tilde H(y,r=0)-1$ approaches the constant value of 0. Comparing the numerical solutions also shows that the metric function approaches asymptotically to the constant value of 1 at shorter distance $r$, and longer distance in $y$-direction. Moreover, compared to the solutions (\ref{GENFFIN}) presented in figure \ref{fig:HM2}, the solutions (\ref{GENfinaltilde}) are approaching to the asymptotic values at longer distances in $r$ and $y$ directions.
}

\begin{figure}[H]
\centering
\includegraphics[width=0.4\textwidth]{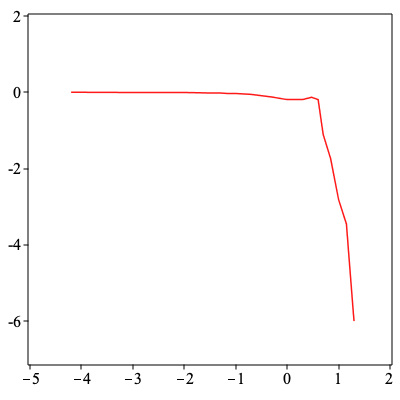}\includegraphics[width=0.4\textwidth]{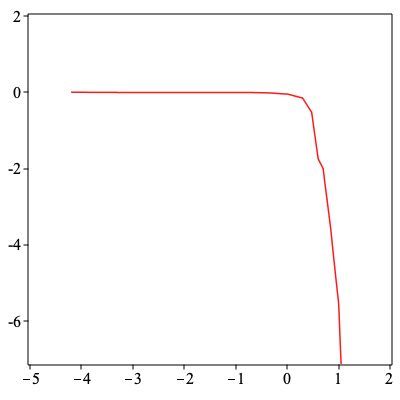}
\caption{The numerical solutions for the logarithm of  $\tilde H(y=0,r)-1$ versus logarithm of $\frac{r}{N}$ (left) and the logarithm of  $\tilde H(y,r=0)-1$ versus logarithm of $\frac{y}{N}$, where we set $N=2$.}
\label{fig:HMM2}
\end{figure}

We can compactify the second M2 brane solution either on $x$ or $z$ directions, with period $2\pi$. We find that the for the Killing vector $\partial / \partial x$, the ten-dimensional metric, is the same as (\ref{ds10}), where $H$ is given by (\ref{GENfinaltilde}). The ten-dimensional metric describes a fully localized D2 brane along the world-volume of a a D6 brane in type IIA string theory. Moreover, the other fields in type IIA theory are still the same as (\ref{dilIIA}), (\ref{CIIA}) and (\ref{A3}), where $H$ is given by (\ref{GENfinaltilde}).
 
We have explicitly verified that the metric (\ref{ds10}) and different fields (\ref{ds10})-(\ref{A3}), satisfy exactly all the ten-dimensional type IIA supergravity field equations, where $H$ is given by (\ref{GENfinaltilde}). Similar to what was presented in section \ref{sec:M2}, the number of preserved supercharges for the membrane solutions with the metric function (\ref{GENfinaltilde}), can be obtained by finding the non-trivial solutions to the Killing spinor equation (\ref{K}). Using equation (\ref{K}) with $M=t,x_1$ and $x_2$, leads to the projection equation (\ref{first}), which eliminates half of the components of the Killing spinor $\epsilon$. Using equation (\ref{K}) with $M=y$ and $M=r$, yields two equations which are trivially satisfied. The equation (\ref{K}) with $M=\alpha_i,\,i=1,2,3$, leads to three equation, as (\ref{K})-(\ref{K3}) with the solution (\ref{eprime}). Moreover, we find  equation (\ref{K}) with $M=z,\theta$ and $x$, leads to the second projection equation (\ref{epp}), which eliminates half of the components of the Killing spinor $\epsilon'$. Hence the number of preserved supersymmetries is eight for the membrane solution with the metric function (\ref{GENfinaltilde}). 
{\textcolor{black}{As we notice from the explicit derivation of the Killing spinor equations for the M2-brane (in appendix \ref{app}), the final projection equations (\ref{first}) and (\ref{epp}) are independent of the explicit dependence of the metric function $H(y,r)$ on the coordinates $y$ and $r$, as given by (\ref{GENfinaltilde}).}}
\section{Embedding the Nutku geometry in M5 brane solutions}
\label{sec:M5}

In this section, we consider the eleven dimensional M5 brane metric, which is given by the line element
\begin{eqnarray}
ds_{11}^{2}&=&H(y,r)^{-1/3}\left(
-dt^{2}+dx_{1}^{2}+dx_{2}^{2}+dx_{3}^{2}+dx_{4}^{2}+dx_{5}^{2}\right) 
+H(y,r)^{2/3}\left( dy^{2}+ds_{Nutku}^{2}\right), \nonumber\\
&&
\label{ds11m5p}
\end{eqnarray}
where the transverse space consists of the Nutku geometry (\ref{Nutku}) with $\epsilon=1$. Moreover, the 
field strength tensor is given by%
\begin{eqnarray}
F_{y\theta xz}&=&-\frac{1}{2}\alpha \sqrt{r^2+N^2}\frac{\partial H}{\partial r}, \label{FScompo1}\\ 
F_{r\theta xz}&=&\frac{1}{2}\alpha r \frac{\partial H}{%
\partial y}.
\label{FScompo2}
\end{eqnarray}%
In equations (\ref{FScompo1}) and (\ref{FScompo2}), $\alpha$ is equal to $1$ for an M5 brane and $-1$ for an  anti M5 brane.

The metric (\ref{ds11m5p}) and the field strength tensor (\ref{FScompo1}) and (\ref{FScompo2}), are solutions to the eleven dimensional
supergravity equations (\ref{GminGG}) and (\ref{dF}), where the metric function  $H\left( y,r\right) $ satisfies
\be
r \sqrt {{a}^{2}+{r}^{2}} {\frac {\partial ^{2}}{\partial {y}^{2}}}H \left( y,r \right) 
  + \left( {a}^{2}+{r}^{2} \right) {
\frac {\partial ^{2}}{\partial {r}^{2}}}H \left( y,r \right) + r
{\frac {\partial }{\partial r}}H \left( y,r \right)=0.\label{PDE5}
\ee
To solve (\ref{PDE5}), we consider 
\begin{equation}
H(y,r)=1+Q_{M5}R(r)Y(y),
\end{equation}%
where $Q_{M5}$ is the charge on the M5-brane. The equation (\ref{PDE5}) then separates to two ordinary differential equations for $R(r)$ and $Y(y)$, respectively. The solutions to 
differential equation for $Y(y)$ are
\be
Y(y)=y_1\sin(cy)+y_2\cos(cy),
\ee
and the solutions to differential equation for $R(r)$, are exactly the same as (\ref{Rsol}).

Furnished with the solutions to equation (\ref{PDE5}) for any separation constant $c$, we find the most general solution to (\ref{PDE5}), by superimposing all solutions with different $c$, as
\begin{eqnarray}
H(y,r)&=&1+Q_{M5}\int _0 ^\infty dc \{ {\cal H}_D(0,0,- N^2c^2,0,\frac{\sqrt{ N^2+r^2}}{r})\times(g_1(c)\sin(cy)+g_2(c)\cos(cy))\nonumber\\
&+&{\cal H}_D (0,-\frac{c^2(N^4-1)}{4},\frac{c^2(N^4+1)}{2},-\frac{c^2(N^4-1)}{4},\frac{N^4+4r\sqrt{r^2+ N^2}-1}{N^4-2 N^2-4r^2+1})\nonumber\\
&\times&(g_3(c)\sin(cy)+g_4(c)\cos(cy))\},\label{GEN5}
\end{eqnarray}
where $g_i(c)$ with $i=1,\cdots ,4$ are four arbitrary functions of the separation constant. We find the functions $g_i(c)$ by considering the near horizon limit. In the limit of $N \rightarrow 0$, the four-dimensional metric (\ref{Nutku}) reduces to $D^2 \times R^2$, where $D^2$ is a two-dimensional disk. In this limit, we find an exact solution to the field equation (\ref{GminGG}) and (\ref{dF}),  as
\begin{eqnarray}
{\widetilde{ds}}_{11}^{2}&=&\tilde H_0(y,r,x,z)^{-2/3}\left(  -dt^{2}+dx_{1}^{2}+dx_{2}^{2}+dx_{3}^{2}+dx_{4}^{2}+dx_{5}^{2}\right)\nn\\
&+&\tilde H_0(y,r,x,z)^{1/3}\left(  dy^2+ds_{D^2 \times R^2}^{2}
\right),  \label{ds11flatM5}
\end{eqnarray}
where 
\be
\tilde H_0(y,r,x,z)=1+\frac{Q_{M5}}{(r^2+x^2+y^2+z^2)^{\frac{3}{2}}}.\label{H05}
\ee

The non-zero components of the four-form field strength are still given by (\ref{FScompo1}) and (\ref{FScompo2}), with replacement $H$ by $\tilde H_0$. We require the metric function (\ref{GEN5}) reduces to (\ref{H05}), in the appropriate near horizon limit, where the M5-brane is located at $x=z=0$. We find an integral equation for $g_i(c)$ ($i=1,\cdots ,4$), which is given by
\begin{eqnarray}
&\lim_{N\rightarrow 0}&\int _0 ^\infty dc \{ {\cal H}_D(0,0,- N^2c^2,0,\frac{\sqrt{ N^2+r^2}}{r})\times(g_1(c)\sin(cy)+g_2(c)\cos(cy))\nonumber\\
&+&{\cal H}_D (0,-\frac{c^2(N^4-1)}{4},\frac{c^2(N^4+1)}{2},-\frac{c^2(N^4-1)}{4},\frac{N^4+4r\sqrt{r^2+ N^2}-1}{N^4-2 N^2-4r^2+1})\nonumber\\
&\times&(g_3(c)\sin(cy)+f_4(c)\cos(cy))\}=\frac{1}{(r^2+y^2)^{\frac{3}{2}}}.\label{INT5}
\end{eqnarray}
To solve (\ref{INT5}) and find $g_i(c)$, we notice the limits of the Heun-D functions can be obtained by looking at the solutions to the radial equation (\ref{Req}) where $N \rightarrow 0$. In fact, we find the limit of the first Heun-D function in (\ref{INT5}), is proportional to the Bessel function $I_0(cr)$, while the limit of the second Heun-D function in (\ref{INT5}), is proportional to the Bessel function $K_0(cr)$. We then find the unique solutions to the integral equation (\ref{INT5}), which are given by
\be
g_1(c)=0,\,g_2(c)=0,\,g_3(c)=0,\,g_4(c)=\frac{2}{\pi} c^2.\label{gs5}
\ee
To summarize, the metric function for the M5 brane solution (\ref{ds11m5p}) is 
\begin{eqnarray}
H(y,r)&=&1+\frac{2Q_{M5}}{\pi}\int _0 ^\infty dc \,c^2 \,\cos(cy)\,  {\cal H}_D (0,-\alpha{c^2},(1+2\alpha)c^2,-\alpha c^2,\frac{4\alpha+4r\sqrt{r^2+ N^2}}{2(1+2\alpha-2r^2-N^2)})
.\nonumber\\
&&\label{GENFFIN5}
\end{eqnarray}
We note that in asymptotic limits, where $y \rightarrow \infty$, or $r \rightarrow \infty$, or both, the integrand is a superposition of a finite oscillating cosine function, and a decaying oscillating Heun-D function.  We numerically evaluate the asymptotic values of the integral in  (\ref{GENFFIN5}). The results show that the decaying Heun-D function takes over the finite oscillating cosine function, and the metric function $H(y,r)$ asymptotically approaches the constant value of 1.
{\textcolor{black}{Although we can't find an analytic expression for the integral in (\ref{GENFFIN5}), we numerically find the behaviour of the metric function, which is shown in figure \ref{fig:HM5}. The left figure in \ref{fig:HM5} shows the logarithm of  $H(y=0,r)-1$ versus logarithm of $\frac{r}{N}$. Moreover, the right figure in \ref{fig:HM5} shows the logarithm of  $H(y,r=0)-1$ versus logarithm of $\frac{y}{N}$. We note that in the extremal asymptotic limit, where  $r \rightarrow \infty$ or  $y \rightarrow \infty$, the logarithm of $H(y=0,r)-1$ or $ H(y,r=0)-1$ approaches $-\infty$, hence the metric function asymptotically approaches the constant value of 1.  On the other extremal limit, i.e. in the near core limit where $r \rightarrow 0$ and $y \rightarrow 0$, the logarithm of $H(y=0,r)-1$ or $H(y,r=0)-1$ approaches the constant value of 0. Comparing the numerical solutions also shows that the metric function approaches asymptotically to the constant value of 1 at shorter distance $r$, and longer distance in  $y$-direction.}

\begin{figure}[H]
\centering
\includegraphics[width=0.4\textwidth]{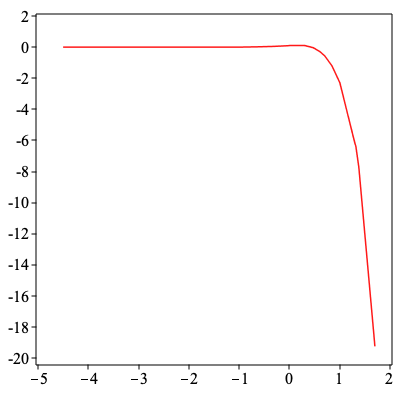}\includegraphics[width=0.4\textwidth]{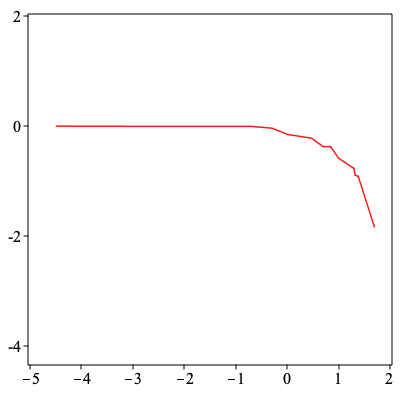}
\caption{The numerical solutions for the logarithm of  $H(y=0,r)-1$ versus logarithm of $\frac{r}{N}$ (left) and the logarithm of  $H(y,r=0)-1$ versus logarithm of $\frac{y}{N}$, where we set $N=2$.}
\label{fig:HM5}
\end{figure}

Considering either the $x$ or $z$ directions to be periodic with period $2\pi$, we find the different fields in ten dimensions by using either the Killing vector $\partial / \partial x$ or  $\partial / \partial z$. For the Killing vector $\partial / \partial x$, we find the ten dimensional metric, which is given by
\begin{eqnarray}
ds_{10}^2&=&\frac{1}{(1+\frac{N^2}{r^2})^{1/4}}\sqrt{{1+\frac{N^2\sin ^2\theta}{r^2}}}\,(-dt^2+dx_1^2+dx_2^2+dx_3^2+dx_4^2+dx_5^2)\nn\\
&+&\frac{H}{(1+\frac{N^2}{r^2})^{1/4}}\sqrt{{1+\frac{N^2\sin ^2\theta}{r^2}}}\,dy^2+\nonumber\\
&+&\frac{H\sqrt{{1+\frac{N^2\sin ^2\theta}{r^2}}}}{(1+\frac{N^2}{r^2})^{3/4}}\,\left({{d{{r}}}^{2}+ \left({N}^{2}+{r}^{2} \right) {d{{\theta}}}^{2}+
 \left( 1+{\frac {{N}^{2}
\cos ^2 \theta}{{r}^{2}}} \right) {d
{{z}}}^{2}
}\right).\label{ds10M5}
\end{eqnarray}
The metric (\ref{ds10M5}) shows an NS5 brane which is fully localized in the world-volume of a D6 brane.  
We find the NSNS dilaton is given by
\be
\Phi=\frac{3}{4}\ln \{\frac{(1+\frac{N^2\sin^2\theta}{r^2})H^{2/3}}{\sqrt{1+\frac{N^2}{r^2}}}\},\label{dilAM5}
\ee
while the NSNS field strength 3-form ${\cal H}_3$ for the associated two-form of the NS5 brane is
\begin{eqnarray}
{\cal H}_{y\theta z}&=&F_{y\theta zx},\\
 {\cal H}_{r\theta z}&=&F_{r\theta zx}.
\end{eqnarray}
The RR fields associated to D6 brane, are given by
\be
C_z=-\frac{N^2\sin \theta \cos\theta}{r^2+{N^2\sin^2\theta}},\label{CIIAM5}
\ee
and
\be
A_{tx_1x_2}=\frac{1}{H}.\label{A3M5}
\ee
We note that in equations (\ref{ds10M5})-(\ref{A3M5}), $H$ is given  by (\ref{GENFFIN5}). We have explicitly verified that the metric (\ref{ds10M5}) and different fields (\ref{dilAM5})-(\ref{A3M5}), satisfy exactly all the ten-dimensional type IIA supergravity field equations.  
We find the number of preserved supersymmetries by finding the non-trivial solutions to the Killing spinor equation (\ref{K}). 
{\textcolor{black}{In appendix \ref{app5}, we present the explicit dependence of $e_{\mu M}$, $\omega_{\mu\nu M}$, $\omega^\mu_{\nu \rho}$ on the metric function $H$ and its derivatives.} By a very similar calculation in section \ref{sec:M2}, we find the projection equation
\be
\Gamma^{\hat t\hat x_1\hat x_2 \hat x_3\hat x_4\hat x_5}\epsilon=\epsilon,\label{first5}
\ee
which eliminates half of the components of the spinor $\epsilon$. {\textcolor{black}{In appendix \ref{app5}, we present explicitly all the different terms in equation (\ref{K}) with $M=t,x_1,\cdots ,x_5$, which leads to equation (\ref{first5}).}} The other remaining Killing spinor equations are
\begin{eqnarray}
\partial _\theta \epsilon&=&(Z_{\hat x \hat z}\Gamma ^{\hat x \hat z}+Z_{\hat \theta \hat r}\Gamma ^{\hat \theta \hat r})\epsilon,\label{M5theta}\\
\partial _x\epsilon&=&(Z'_{\hat r \hat x}\Gamma ^{\hat r \hat x}+Z'_{\hat x \hat\theta}\Gamma ^{\hat x\hat\theta }+Z'_{\hat r \hat z}\Gamma ^{\hat r \hat z}+Z'_{\hat z \hat \theta}\Gamma ^{\hat z\hat \theta})\epsilon,\label{M5x}\\
\partial _{z}\epsilon&=&(Z_{\hat z \hat r} \Gamma^{\hat z\hat r}+Z_{\hat \theta \hat x} \Gamma^{\hat \theta \hat x}+Z_{\hat \theta \hat z} \Gamma^{\hat \theta \hat z}+Z_{\hat x \hat r} \Gamma^{\hat x \hat r})\epsilon,\label{K5M5}
\end{eqnarray}
where all the coefficients are given in appendix \ref{app.gamma}. {\textcolor{black}{We present derivation of equations (\ref{M5theta})-(\ref{K5M5}) in appendix \ref{app5}.}} We find the projection equation
\be
\Gamma^{\hat x\hat z\hat r\hat \theta}\epsilon=\epsilon,\label{secK}
\ee
solves equations (\ref{M5theta})-(\ref{K5M5}) up to a rotational transformation. Equation (\ref{secK}) eliminates another half of the components of the spinor $\epsilon$. So, the M5 brane solution (\ref{ds11m5p}) preserves eight supercharges.  {\textcolor{black}{As we notice from the explicit derivation of the Killing spinor equations for the M5-brane (in appendix \ref{app5}), the final projection equations (\ref{first5}) and (\ref{secK}) are independent of the explicit dependence of the metric function $H(y,r)$ on the coordinates $y$ and $r$, as given by (\ref{GENFFIN5}).}}

We can find type IIB intersecting NS5 brane with D5, by applying the T-duality on IIA solutions (\ref{ds10M5}). Applying T-duality in $x_1$-direction gives
\begin{eqnarray}
\widetilde{ds}_{10}^2&=&\frac{1}{(1+\frac{N^2}{r^2})^{1/4}}\sqrt{{1+\frac{N^2\sin ^2\theta}{r^2}}}\,(-dt^2+\frac{(1+\frac{N^2}{r^2})^{1/2}}{{{1+\frac{N^2\sin ^2\theta}{r^2}}}}dx_1^2+dx_2^2+dx_3^2+dx_4^2+dx_5^2)\nn\\
&+&\frac{H}{(1+\frac{N^2}{r^2})^{1/4}}\sqrt{{1+\frac{N^2\sin ^2\theta}{r^2}}}\,dy^2+\nonumber\\
&+&\frac{H\sqrt{{1+\frac{N^2\sin ^2\theta}{r^2}}}}{(1+\frac{N^2}{r^2})^{3/4}}\,\left({{d{{r}}}^{2}+ \left({N}^{2}+{r}^{2} \right) {d{{\theta}}}^{2}+
 \left( 1+{\frac {{N}^{2}
\cos ^2 \theta}{{r}^{2}}} \right) {d
{{z}}}^{2}
}\right),\label{ds10M5IIB}
\end{eqnarray}
which describes type IIB NS5 brane localized on 4 directions of a D5 brane. We note that type IIB dilaton field is
\be
\tilde \Phi=\frac{1}{2}\ln \{\frac{(1+\frac{N^2\sin^2\theta}{r^2})H}{\sqrt{1+\frac{N^2}{r^2}}}\},\label{dilAM5IIB}
\ee
and the components of the dual NSNS two form are
\begin{eqnarray}
\tilde B_{rz}&=&B_{rz},\label{B1}\\
\tilde B_{yz}&=&B_{yz}.\label{B2}
\end{eqnarray}
The IIB RR axion field and four-form field are identically zero, while the only non-zero component of the RR two form field is $\tilde {\cal B}_{zx_1}=C_z$. The charge of D5 brane is the integral of $\tilde {\cal B}_{zx_1}$ over an $S^3$. We have explicitly verified that the metric (\ref{ds10M5IIB}), together with the other fields (\ref{dilAM5IIB})-(\ref{B2}) and the RR two form, satisfy all the field equations of the ten-dimensional type IIB supergravity. Moreover, we have explicitly solved the Killing spinor equation (\ref{K}), and verified the solution (\ref{ds10M5IIB}), preserves eight supersymmetries.
\section{Decoupling limits }
\label{sec:DL}

In this section, we discuss the decoupling of the D2 or NS5 branes from the bulk brane, at low energies. First we consider the D brane system (\ref{ds10}). At low energy, the region near to the D6 brane is governed by the energy scale of the infrared fixed point. We also get massless fundamental hyper-multiplets  in the corresponding field theory of the D2 brane. The field theory limit, near the D2 horizon is given by $g^2_{YM2}=\frac{g_s}{l_s}$, where $g_s$ and $l_s$, are the string coupling constant and length scale, respectively, and $g_{YM2}$ is the coupling constant of Yang-Mills theory on the world-volume of D2 brane. In the limit, the gauge theory coupling on $D6$ brane $g_{YM6} \propto  g_{YM2}l_s^2$, approaches zero, and so we find the decoupling in the D brane system. We can rescale the coordinates $y,r$, according to $y=Yl_s^2$, and $r=Wl_s^2$, such that $Y$ and $W$ are fixed quantities. The metric function (\ref{GENFFIN}) scales as
\be
H(Y,W)=\frac{h(Y,W)}{l_s^4},
\ee
where 
\begin{eqnarray}
h(Y,W)&=&-\frac{32\pi^2N_2g_{YM2}^4}{Y}\int _0 ^\infty dC C^4 {J_1(CY)}  {\cal H}_D (0,-A{C^2},2AC^2,-AC^2,\frac{2A+2W\sqrt{W^2+ {\cal N}^2}}{2A-2W^2-{\cal N}^2})
,\nonumber\\
&&\label{GENFFINscaled}
\end{eqnarray}
is clearly independent of $l_s$, and shows the decoupling of D2 brane from the bulk of D brane system. The rescaled quantities $C$, $A$ and ${\cal N}$ in (\ref{GENFFINscaled}), are given by
\begin{eqnarray}
C&=&cl_s^2,\\
A&=&\frac{\alpha}{l_s^4},\\
\cal N&=&\frac{N}{l_s^2},
\end{eqnarray}
and we use the relation $Q_{M2}=32\pi^2N_2g_s^2l_s^6$, for the charge of M2 brane, where $N_2$ is the number of D2 branes. We also notice that in the decoupling limit, the ten-dimensional metric (\ref{ds10}), for the decoupled D brane system, becomes
\begin{eqnarray}
\frac{ds_{10}^2}{l_s^2}&=&\frac{h(Y,W)^{-1/2}}{(1+\frac{{\cal N}^2}{W^2})^{1/4}}\sqrt{{1+\frac{{\cal N}^2\sin ^2\theta}{r^2}}}(-dt^2+dx_1^2+dx_2^2)+\frac{h(Y,W)^{1/2}}{(1+\frac{{\cal N}^2}{r^2})^{1/4}}\sqrt{{1+\frac{{\cal N}^2\sin ^2\theta}{r^2}}}ds_4^2(Y)\nonumber\\
&+&\frac{h(Y,W)^{1/2}}{(1+\frac{{\cal N}^2}{W^2})^{3/4}}\sqrt{{1+\frac{{\cal N}^2\sin ^2\theta}{W^2}}}({{d{{W}}}^{2}+ \left({\cal N}^{2}+{W}^{2} \right) {d{{\theta}}}^{2}+
 \left( 1+{\frac {{\cal N}^{2}
\cos ^2 \theta}{{W}^{2}}} \right) {d
{{Z}}}^{2}
}),\label{ds10dec}
\end{eqnarray}
where $Z=\frac{z}{l_s^2}$. The metric depends on $l_s$ only by an overall factor, which indicates the supergravity dual of a quantum field theory. 

Now, we consider the decoupling limit for the fully localized NS5 brane in the world volume of D6, as given by the type IIA solution (\ref{ds10M5}). Similar to what we had for the D2 brane, the NS5 decouples from the bulk, at low energy. So, we consider the limit where $g_s \rightarrow 0$ with a fix $l_s$. We rescale the coordinates  $y,r$, according to $y=Yg_sl_s^2$, and $r=Wg_sl_s^2$, such that $Y$ and $W$ are fixed quantities. We find that the metric function (\ref{GENFFIN5}) scales as 
\be
H(Y,W)=\frac{h(Y,W)}{g_s^2},
\ee
where
\begin{eqnarray}
h(Y,W)&=&\frac{2N_5}{l_s^3}\int _0 ^\infty dC \,C^2 \,\cos(CY)\,  {\cal H}_D (0,-A{C^2},2AC^2,-AC^2,\frac{2A+2W\sqrt{W^2+{\cal N}^2}}{2A-2W^2-{\cal N}^2})
,\nonumber\\
&&\label{GENFFIN5DEC}
\end{eqnarray}
is clearly independent of $g_s$, and shows the decoupling of NS5 brane from the bulk of the brane system. Note that in (\ref{GENFFIN5DEC}), we rescale $c,\alpha$ and $N$ to
\begin{eqnarray}
C&=&cg_sl_s^2,\label{cr}\\
A&=&\frac{\alpha}{g_s^2l_s^4},\\
\cal N&=&\frac{N}{g_sl_s^2},\label{Nr}
\end{eqnarray}
and use the relation $Q_{M5}=\pi N_5 l_p^3$ for the charge of M5 brane, where $N_5$ is the number of NS5 branes. We also notice that in the decoupling limit, the ten-dimensional metric (\ref{ds10M5}) for the decoupled NS-D brane system, becomes
\begin{eqnarray}
ds_{10}^2&=&\frac{1}{(1+\frac{{\cal N}^2}{W^2})^{1/4}}\sqrt{{1+\frac{{\cal N}^2\sin ^2\theta}{W^2}}}\,(-dt^2+dx_1^2+dx_2^2+dx_2^2+dx_3^2)\nn\\
&+&\frac{h(Y,W)l_s^4}{(1+\frac{{\cal N}^2}{W^2})^{1/4}}\sqrt{{1+\frac{{\cal N}^2\sin ^2\theta}{W^2}}}\,dY^2+\nonumber\\
&+&\frac{h(Y,W)l_s^4\sqrt{{1+\frac{{\cal N}^2\sin ^2\theta}{W^2}}}}{(1+\frac{{\cal N}^2}{W^2})^{3/4}}\,\left({{d{{W}}}^{2}+ \left({\cal N}^{2}+{W}^{2} \right) {d{{\theta}}}^{2}+
 \left( 1+{\frac {{{\cal N}}^{2}
\cos ^2 \theta}{{W}^{2}}} \right) {d
{{Z}}}^{2}
}\right),\nn\\
&&\label{ds10M5dec}
\end{eqnarray}
where we rescale $z$ to $z=Zg_sl_s^2$.  The decoupled free theory on the world-volume of NS5, is the non-gravitational little string theory \cite{shiraz}, in which the modes of the theory self-interact, and are decoupled from the bulk. We also note that the little string theory possesses the same T-duality as the type IIA supergravity, as taking the limit $g_s \rightarrow 0$, commutes with the T-duality transformation. We also may apply he T-duality transformation on the compactified little string theory. In case of toroidal compactification on a $d$-torus, the T-transformation is homomorphic to $O(d,d,{\mathbb Z})$. 

We also can find the decoupling limit for the type IIB supergravity solution (\ref{ds10M5IIB}), where the NS5 brane is partially localized in the world volume of D5 brane. The field theory limit, in which NS5 decouples from the bulk at low energy, is given by fixed values for  $g_{YM5}=l_s$, where  $g_{YM5}$ is the coupling constant for the Yang-Mills theory on the world-volume of NS5 brane. Similar to type IIA system, we rescale the coordinates  $y,r$, according to $y=Yg_sl_s^2$, and $r=Wg_sl_s^2$, such that $Y$ and $W$ are fixed quantities. The metric function (\ref{GENFFIN5}) scales as 
\be
H(Y,W)=\frac{h(Y,W)}{g_s^2},
\ee
where 
\begin{eqnarray}
h(Y,W)&=&\frac{2N_5}{g^3_{YM5}}\int _0 ^\infty dC \,C^2 \,\cos(CY)\,  {\cal H}_D (0,-A{C^2},2AC^2,-AC^2,\frac{2A+2W\sqrt{W^2+{\cal N}^2}}{2A-2W^2-{\cal N}^2})
,\nonumber\\
&&\label{GENFFIN5DECIIB}
\end{eqnarray}
where we rescale $c,\alpha$ and $N$, according to (\ref{cr})-(\ref{Nr}), respectively. We also notice that in the decoupling limit, the ten-dimensional metric (\ref{ds10M5}) for the decoupled NS5 and D5 branes, becomes
\begin{eqnarray}
\widetilde{ds}_{10}^2&=&\frac{1}{(1+\frac{{\cal N}^2}{W^2})^{1/4}}\sqrt{{1+\frac{{\cal N}^2\sin ^2\theta}{W^2}}}\,(-dt^2+\frac{(1+\frac{{\cal N}^2}{W^2})^{1/2}}{{{1+\frac{{\cal N}^2\sin ^2\theta}{W^2}}}}dx_1^2+dx_2^2+dx_3^2+dx_4^2+dx_5^2)\nn\\
&+&\frac{g_{YM5}^4h(Y,W)}{(1+\frac{{\cal N}^2}{W^2})^{1/4}}\sqrt{{1+\frac{{\cal N}^2\sin ^2\theta}{W^2}}}\,dY^2+\nonumber\\
&+&\frac{g_{YM5}^4h(Y,W)\sqrt{{1+\frac{{\cal N}^2\sin ^2\theta}{W^2}}}}{(1+\frac{{\cal N}^2}{W^2})^{3/4}}\,\left({{d{{W}}}^{2}+ \left({\cal N}^{2}+{W}^{2} \right) {d{{\theta}}}^{2}+
 \left( 1+{\frac {{\cal N}^{2}
\cos ^2 \theta}{{r}^{2}}} \right) {d
{{Z}}}^{2}
}\right),\nn\\
&&\label{ds10M5IIBdec}
\end{eqnarray}
where we rescale $z$ to $z=Zg_sl_s^2$. We notice from (\ref{ds10M5IIBdec}) that the low energy limit of the decoupled theory is a Yang-Mills theory, where the coupling constant is equal to the string length scale. Moreover, in the limit of $g_s \rightarrow 0$, the decoupled free theory on the NS5 brane becomes a type IIB $(1,1)$ little string theory, which possesses eight supercharges \cite{AHA}. The preserved number of supersymmetries is in perfect agreement with what we found for the number of preserved supersymmetries for the ten-dimensional type IIB supergravity solution (\ref{ds10M5IIB}). Of course, the type IIB supergravity solution (\ref{ds10M5IIB}) preserves eight supersymmetries, as the same as type IIA solution (\ref{ds10M5}), because the former solution is T-dual to the latter. We should mention that for a flat Euclidean  transverse geometry, the system of $N_5$ NS5-branes located at $N_6$ D6-branes is the result of dimensional reduction of $N_5N_6$ coinciding M5-branes. In the limit of $g_s \rightarrow 0$, the world-volume theory of the decoupled IIA NS5-branes, is the non-gravitational six dimensional little string theory \cite{seiberg}. The theory has $(2,0)$ supersymmetry and an R-symmetry, which is the leftover of the ten dimensional Lorentz symmetry $SO(1,9)$. Considering the D6 bulk,  breaks the supersymmetry down to $(1,0)$, with eight supersymmetries. We also find that the IIB supergravity solution (\ref{ds10M5IIB}) possesses eight supersymmetries which indicate the decoupled theory, described by the first line of equation (\ref{ds10M5IIBdec}), is a new little string theory.


\section{Conclusions}

We construct eleven-dimensional supergravity solutions for the M2 and M5 branes. The construction of the solutions is based on uplifting the  Nutku geometry into M-theory.  The solutions are realization of new fully localized type IIA D2 and D6 branes, as well as NS5 and D6 branes. 
The brane metric function for all solutions, is a convoluted integral of two functions.  Dimensional reduction of the solutions, on a compact Killing direction of the Nutku geometry, leads to type IIA brane solutions, which preserve eight supersymmetries. We explicitly drive the number of preserved supersymmetries for each solution. We also find that T-dualizing the system of fully localized IIA NS5 and D6 brane, leads to a type IIB partially localized NS5 and D5 branes. We discuss the decoupling limits of the solutions and find the D2 and NS5 can decouple from the D6 bulk. We also find the decoupled NS5 brane describes a new type IIB little string theory. It would be quite interesting to consider the constructed M-brane solutions, as the holographic dual theory for the NS5 world-volume theory with the matter coming from the D6 branes. The supergravity solutions may be used to calculate the correlation functions, and the spectrum of fields in the new little string theory. As an example, the two point function of the energy-momentum tensor of the little string theory, may be calculated from the on-shell classical action of the supergravity with the field solutions \cite{shiraz}. The holographic duality also may be used to find some states of the new little string theory.

\vskip 1cm
{\Large Acknowledgments}

The author would like to thank M. Butler for a short discussion in the early stage of the work. This work was supported by the Natural Sciences and Engineering Research Council of Canada. All data generated or analyzed during this study are included in this published article.

{\textcolor{black}{
\appendix
\section{The Killing spinor equation for M2-brane}\label{app}
The elfbein components are given by
\begin{eqnarray}
e_{\hat t t}&=&H^{-1/3},\\
e_{\hat x_1 x_1}&=&H^{-1/3},\\
e_{\hat x_2 x_2}&=&H^{-1/3},\\
e_{\hat y y}&=&H^{1/6},\\
e_{\hat \alpha_1 \alpha_1}&=&yH^{1/6},\\
e_{\hat \alpha_2 \alpha_2}&=&y\sin(\alpha_1)H^{1/6},\\
e_{\hat \alpha_3 \alpha_3}&=&y\sin(\alpha_1)\sin(\alpha_2)H^{1/6},\\
e_{\hat z r}&=&f^{-1/4}H^{1/6},\\
e_{\hat r \theta}&=&rf^{1/4}H^{1/6},\\
e_{\hat \theta x}&=&\frac{rf^{1/4}H^{1/6}}{(r^2+N^2\cos^2\theta)^{1/2}},\\
e_{\hat x z}&=&\frac{(r^2+N^2\cos^2\theta)^{1/2}H^{1/6}}{rf^{1/4}},\\
e_{\hat x x}&=&\frac{N^2\sin\theta \cos\theta H^{1/6}}{f^{1/4}(r^2+N^2\cos^2\theta)^{1/2}},
\end{eqnarray}
where 
\be
f\equiv f(r)=(1+\frac{N^2}{r^2}).\label{ffrr}
\ee
We then find the components of tensor $h^\mu _{\nu \rho}$, according to equation (\ref{de}), which in turn give the component of tensor $\omega_{\mu \nu \rho}$, as in equation (\ref{omeg}).  We list here few components of  $\omega_{\mu \nu \rho}$ as there are 32 non-zero components.
\begin{eqnarray}
\omega_{\hat t \hat y \hat t}&=&\frac{H'_y}{3H^{7/6}},\label{o1}\\
\omega_{\hat t \hat z \hat t}&=&\frac{f^{1/4}H'_r}{3H^{7/6}},\\
\omega_{\hat y\hat \alpha_1 \hat \alpha_1}&=-&\frac{6H+yH'_y}{6yH^{7/6}},\\
\omega_{\hat \theta\hat x\hat r}&=-&\frac{N^2(N^2\cos^2\theta+2r^2\cos^2\theta-r^2)}{2f^{3/4}r^{5/2}H^{1/6}(r^2+N^2\cos^2\theta)},\\
\omega_{\hat r\hat x \hat x}&=&\frac{N^2\sin\theta\cos\theta}{f^{1/4}r(r^2+N^2\cos^2\theta))H^{1/6}},\\
\omega_{\hat z\hat r\hat r}&=-&\frac{ra^2H'_r+r^3H'_r+3N^2H+6r^2H}{6f^{3/4}r^3H^{7/6}}.
\end{eqnarray}
We find now the components of $\omega_{M \nu \rho}$  from equation (\ref{omeg2}). There are 65 non-zero components. We list here a few non-zero  $\omega_{M \nu \rho}$, which are given by
\begin{eqnarray}
\omega_{ t\hat y\hat t}&=&-\frac{H'_y}{3H^{3/2}},\\
\omega_{ t \hat z \hat t}&=&-\frac{f^{1/4}H'_r}{3H^{3/2}},\\
\omega_{y\hat \alpha_1 \hat \alpha_1}&=&-\frac{6H+yH'_y}{6yH},\\
\omega_{ \theta\hat \theta \hat x}&=&\frac{N^2(N^2\cos^2\theta+2r^2\cos^2\theta-r^2)}{2f^{1/2}r^2(r^2+N^2\cos^2\theta)},\\
\omega_{ r\hat x \hat x}&=&\frac{(-N^4r\cos^2\theta-N^2r^3\cos^2\theta-N^2r^3-r^5)H'_r+(3N^4\cos^2\theta+6N^2r^2\cos^2\theta-3N^2r^2)H}{6fr^3(r^2+N^2\cos^2\theta))H},\nonumber\\
&&\\
\omega_{z\hat r\hat \theta}&=&-\frac{N^2(N^2\cos^2\theta+2r^2\cos^2\theta-r^2)}{2fr^4(r^2+N^2\cos^2\theta)^{1/2}}.\label{o12}
\end{eqnarray}
We note that in (\ref{o1})-(\ref{o12}), $H'_r=\frac{\partial H}{\partial r}$ and $H'_y=\frac{\partial H}{\partial y}$. 
The third term in the Killing spinor equation (\ref{K}) is
\be
\Gamma _M^{NPQR}F_{NPQR}=\Gamma_M^{tx_1x_2y}F_{tx_1x_2y}+\Gamma_M^{tx_1x_2r}F_{tx_1x_2r}=\frac{-1}{2H^2}(\Gamma_M^{tx_1x_2y}H'_y+\Gamma_M^{tx_1x_2r}H'r),
\ee
where $\Gamma_M^{tx_1x_2y}$ and $\Gamma_M^{tx_1x_2r}$ are defined in (\ref{gamn}). 
Similarly the fourth term in the Killing spinor equation (\ref{K}) is given in terms of $\Gamma ^{PQR}$ multiplied by components  $F_{tx_1x_2y}=-\frac{1}{H^2}H'_y$ or $F_{tx_1x_2r}=-\frac{1}{H^2}H'_r$.
Using the octonionic representation (\ref{oct1})-(\ref{oct4}), we find the components of $\Gamma_M^{tx_1x_2y}$, $\Gamma_M^{tx_1x_2r}$ and $\Gamma ^{PQR}$. Collecting all the terms in the Killing spinor equation (\ref{K}) with $M=t$, and assuming $\partial _t \epsilon=0$, we find
\be
{\cal T}_1(\Gamma_{\hat t \hat y}+\Gamma_{\hat x_1\hat x_2 \hat y})\epsilon+{\cal T}_2(\Gamma_{\hat r\hat t}+\Gamma_{\hat r\hat x_1\hat x_2})\epsilon=0,\label{Kt}
\ee
where
\be
{\cal T}_1=-\frac{1}{6H^{9/6}(y,r)}\left( {\frac {
\partial }{\partial y}}H \left( y,r \right)  \right),
\ee
and
\be
{\cal T}_2=\frac {(r^2+{a}
^{2})^{1/4}}{6 r^{3/4}H^{9/6}(y,r)   }  \left( {\frac {
\partial }{\partial r}}H \left( y,r \right)  \right).
\ee
Quite interestingly, we notice the dependence of the Killing spinor equation to the metric function and its derivative are contained in two independent functions ${\cal T}_1$ and ${\cal T}_2$. Multiplying equation (\ref{Kt}) from left by $\Gamma_{\hat t}$ and from right by $\Gamma_{\hat y}$ leads to 
\be
(\Gamma_{\hat t\hat x_1 \hat x_2}+1)\epsilon=0,\label{firstt}
\ee
which is exactly the equation (\ref{first}). Similar calculations show the Killing spinor equation (\ref{K}) with $M=x_1$ and $M=x_2$ lead to the same equation (\ref{firstt}) or (\ref{first}). Moreover, we notice that projection equation (\ref{firstt}) is independent of the integral equation (\ref{GENFFIN}) for the metric function.  
The Killing spinor equation (\ref{K}) with $M=y$ has two contributions from the second term and third and fourth terms. The contribution from the second term in (\ref{K}) is given by
\be
-\frac{1}{12r^{1/2}H(y,r)}(N^2+r^2)^{1/4}\left(\frac {
\partial }{\partial r}H(y,r)\right)\Gamma_{\hat r\hat y}\epsilon.\label{Ky}
\ee
The contribution from the third and fourth terms in (\ref{K}) is equal to 
\be
-\frac{1}{12r^{1/2}H(y,r)}(N^2+r^2)^{1/4}\left(\frac {
\partial }{\partial r}H(y,r)\right)\Gamma_{\hat y\hat r}\epsilon.\label{Ky2}
\ee
Hence the Killing spinor equation (\ref{K}) with $M=y$, is trivially satisfied. A similar calculation shows that the Killing spinor equation (\ref{K}) with $M=r$ is also trivially satisfied.
The Killing spinor equation (\ref{K}) with $M=\alpha_1$ is
\be
\partial _{\alpha_1}\epsilon+{\cal A}_1(\Gamma_{\hat \alpha_1 \hat y}+\Gamma_{\hat \alpha_1 \hat t\hat x_1\hat x_2 \hat y})\epsilon+{\cal A}_2(\Gamma_{\hat \alpha_1\hat r}+\Gamma_{\hat \alpha_1\hat t\hat x_1\hat x_2\hat r})\epsilon+{\cal A}_3\Gamma_{\hat \alpha_1\hat y}\epsilon=0,\label{Ka1}
\ee
where the three independent functions ${\cal A}_i,\,i=1,2,3$ are given by
\be
{\cal A}_1=\frac{y}{12H(y,r)} \left( {\frac {
\partial }{\partial y}}H \left( y,r \right)  \right),
\ee
\be
{\cal A}_2=\frac{y(a^2+r^2)^{1/4}}{12r^{1/2}H(y,r)} \left( {\frac {
\partial }{\partial r}}H \left( y,r \right)  \right),
\ee
and
\be
{\cal A}_3=\frac{1}{2}.
\ee
Using equation (\ref{firstt}) in (\ref{Ka1}), we find
\be
(\partial _{\alpha_1}+{\cal A}_3\Gamma_{\hat \alpha_1\hat y})\epsilon=0,
\ee
which is exactly equation (\ref{K1}). Similar calculations show the Killing spinor equation (\ref{K}) with $M=\alpha_2$ and $M=\alpha_3$ leads to equations (\ref{K2}) and (\ref{K3}).
We consider the Killing spinor equation (\ref{K}) with $M=z$ now. Solving the Killing spinor equations (\ref{K1})-(\ref{K3}) leads to equation (\ref{eprime}) for the new Killing spinor $\epsilon'$. Using (\ref{eprime}), we find
\be
\partial _{z}\epsilon'+{\cal Z}_1(\Gamma_{\hat z \hat y}+\Gamma_{\hat z \hat t\hat x_1\hat x_2 \hat y})\epsilon'+{\cal Z}_2(\Gamma_{\hat z\hat r}+\Gamma_{\hat z\hat t\hat x_1\hat x_2\hat r})\epsilon'-(Z_{\hat z \hat r} \Gamma^{\hat z\hat r}+Z_{\hat \theta \hat x} \Gamma^{\hat \theta \hat x}+Z_{\hat \theta \hat z} \Gamma^{\hat \theta \hat z}+Z_{\hat x \hat r} \Gamma^{\hat x \hat r})\epsilon'=0,\nonumber
\ee
\be
\label{Kzz}
\ee
where the two independent functions ${\cal Z}_i,\,i=1,2$ are given by
\be
{\cal Z}_1=\frac{1}{12}\,\frac {1}{  H(y,r) }
  \,\sqrt {{\frac {  {N}
^{2}  \cos^{2} \theta  +{r}^{2} 
}{r\sqrt {  {N}^{2}+{r}^{2}}}}} \left( {\frac {\partial }{\partial y}}H \left( y
,r \right)  \right) ,
\ee
and
\be
{\cal Z}_2=\frac{1}{12}\,\frac {1}{ H(y,r){r}}
\sqrt {{ {  {N}^{2}  \cos^{2}\theta
  +{r}^{2} }}}
 \left( {\frac {\partial }{\partial r}}H \left( y,r \right) 
 \right) ,
\ee
and the other $Z$ functions in (\ref{Kzz}) are given in appendix \ref{app.gamma}. We note that ${\cal Z}_1$ and ${\cal Z}_2$ depend on the derivative of the metric function, while the other $Z$ functions do not. Equations (\ref{firstt}) and (\ref{eprime}) lead to
\be
\Gamma ^{\hat t \hat x_1 \hat x_2}\epsilon'=-\epsilon',\label{firstprime}
\ee
which is exactly the equation (\ref{first}).
Using (\ref{firstprime}) in (\ref{Kzz}) gives 
\be
\partial _{z}\epsilon'-(Z_{\hat z \hat r} \Gamma^{\hat z\hat r}+Z_{\hat \theta \hat x} \Gamma^{\hat \theta \hat x}+Z_{\hat \theta \hat z} \Gamma^{\hat \theta \hat z}+Z_{\hat x \hat r} \Gamma^{\hat x \hat r})\epsilon'=0,\label{Kzz11}
\ee
which is exactly equation (\ref{K5}). Multiplying equation (\ref{Kzz11}) from left by $\Gamma^{\hat \theta\hat z}$ and assuming $\epsilon'$ is independent of $z$, leads to
\be
\Gamma^{\hat \theta \hat z \hat x \hat r} \epsilon'=\epsilon',\label{eppAPp2}
\ee
which is exactly the equation (\ref{epp}). Similar calculations show that the Killing spinor equation (\ref{K}) with $M=\theta$ and $M=x$ lead to the same projection equation (\ref{eppAPp2}) for the spinor $\epsilon'$. Each projection equation (\ref{firstprime}) and (\ref{eppAPp2}) breaks the number of supersymmetries by a factor of $\frac{1}{2}$, hence the M2 brane solution preserves $\frac{1}{4}$ of the supersmmetries.
}}
\section{The coefficients of the Killing spinor equations (\ref{K5}) and (\ref{M5theta})-(\ref{K5M5})}\label{app.gamma}
The coefficients which appear in the Killing spinor equations (\ref{K5}) and (\ref{K5M5}), are given by
\begin{eqnarray}
Z_{\hat z \hat r}&=&\,{\frac {{N}^{2} \left( {N}^{2} \cos^2
  \theta  +2\, r^2\cos^2  \theta
 -{r}^{2} \right) }{4\sqrt {{r}^{2}+{N}^{2}
  \cos^2  \theta}\,{r}^{2} \left( {N}^{2
}+{r}^{2} \right) }},  \\
Z_{\hat \theta \hat x}&=&-\,{\frac {{N}^{2} \left( {N}^{2} \cos^2
  \theta  +2\, r^2\cos^2  \theta
 -{r}^{2} \right) }{4\sqrt {{r}^{2}+{N}^{2}
  \cos^2  \theta}\,{r}^{2} \left( {N}^{2
}+{r}^{2} \right) }},   \\
Z_{\hat \theta \hat z} &=&-\,{\frac {{N}^{2}\sin  \theta  \cos \theta
 }{2r\sqrt {{N}^{2}+{r}^{2}}\sqrt {{r}^{2
}+{N}^{2}  \cos^2 \theta }}},\\
Z_{\hat x \hat r} &=&-\,{\frac {{N}^{2}\sin  \theta \cos  \theta
 }{2r\sqrt {{N}^{2}+{r}^{2}}\sqrt {{r}^{2}+{N}
^{2} \cos^2 \theta }}}.
\end{eqnarray}

Moreover, the coefficients $Z_{\hat x \hat z}$ and $Z_{\hat \theta \hat r}$, which appear in equation (\ref{M5theta}), are given by 
\begin{eqnarray}
Z_{\hat x \hat z}&=&{\frac { \left(  \left( {N}^{2}+2\,{r}^{2}
 \right)   \cos ^2 \theta -{r}^{2}
 \right) {N}^{2}}{\sqrt {r} \left( 4\,  \cos ^2 \theta
 \sqrt {{N}^{2}+{r}^{2}}\,{N}^{2}+{r}^{2} \right) },
}\\
Z_{\hat \theta \hat r}&=&-
{\frac { \left(  \left( {N}^{2}+2\,{r}^{2} \right)   \cos^2
 \theta +{r}^{2} \right) {N}^{2}}{\sqrt {r} \left( 4\,  \cos^2 \theta 
 \sqrt {{N}^{2}+{r}^{2}}\,{N}^{2}+{r}^{2} \right) }}.
\end{eqnarray}

The four coefficients which appear in equation (\ref{M5x}), are given by
\begin{eqnarray}
Z'_{\hat r \hat x}&=&-\,{\frac { \left( 2\,  \cos ^2 \theta -1 \right) {N}^{2}}{4r \left( {N}^{2}+{r}^{2}
 \right) ^{3/4}\sqrt [4]{{N}^{2} \cos ^2\theta +{r}^{2}}}},\\
Z'_{\hat x \hat\theta}&=&-\,{\frac {\sin  \theta  \cos \theta  
 \left( {N}^{2}+2\,{r}^{2} \right){N}^{2}}{4{r}^
{2}\sqrt [4]{{N}^{2} \cos ^2\theta +{
r}^{2}} \left( {N}^{2}+{r}^{2} \right) ^{5/4}}},\\
Z'_{\hat r \hat z}&=&\,{\frac {\sin  \theta  \cos \theta  
 \left( {N}^{2}+2\,{r}^{2} \right){N}^{2}}{4{r}^
{2}\sqrt [4]{{N}^{2} \cos ^2\theta +{
r}^{2}} \left( {N}^{2}+{r}^{2} \right) ^{5/4}}},\\
Z'_{\hat z \hat \theta}&=&\,{\frac { \left(  \cos^2  \theta
 -1/2 \right) {N}^{2}}{2r\sqrt [4]{{N}^{2}
\cos ^2\theta +{r}^{2}} \left( {N}^{2}+{r}^{
2} \right) ^{3/4}}}.
\end{eqnarray}
{\textcolor{black}{
\section{The Killing spinor equation for M5-brane}\label{app5}
The elfbein components are given by
\begin{eqnarray}
e_{\hat t t}&=&H^{-1/6},\\
e_{\hat x_1 x_1}&=&H^{-1/6},
\end{eqnarray}
\begin{eqnarray}
e_{\hat x_2 x_2}&=&H^{-1/6},\\
e_{\hat x_3 x_3}&=&H^{-1/6},\\
e_{\hat x_4 x_4}&=&H^{-1/6},\\
e_{\hat x_5 x_5}&=&H^{-1/6},\\
e_{\hat y y}&=&H^{1/3},\\
e_{\hat r r}&=&f^{-1/4}H^{1/3},\\
e_{\hat \theta \theta}&=&rf^{1/4}H^{1/3},\\
e_{\hat x x}&=&\frac{rf^{1/4}}{\sqrt{r^2+N^2\cos^2\theta}}H^{1/3},\\
e_{\hat z x}&=&-\frac{N^2\sin\theta\cos\theta}{f^{1/4}r(r^2+N^2\cos^2\theta)^{1/2}}H^{1/3},\\
e_{\hat z z}&=&\frac{\sqrt{1+\frac{N^2\cos^2\theta}{r^2}}}{f^{1/4}}H^{1/3},
\end{eqnarray}
where $f$ is given by (\ref{ffrr}).
We find the components of tensor $h^\mu _{\nu \rho}$, according to equation (\ref{de}), which in turn give the component of tensor $\omega_{\mu \nu \rho}$, as in equation (\ref{omeg}).  We list here few components of  $\omega_{\mu \nu \rho}$ as there are 28 non-zero components.
\begin{eqnarray}
\omega_{\hat t \hat y \hat t}&=&\frac{H'_y}{6H^{4/3}},\label{o15}\\
\omega_{\hat t \hat r \hat t}&=&\frac{f^{1/4}H'_r}{6H^{4/3}},\\
\omega_{\hat y\hat r\hat r}&=&-\frac{H'_y}{3H^{4/3}},\\
\omega_{\hat r\hat \theta\hat \theta}&=&-{\frac {2\, H_r{N}^{4}r+4\,{a}^{2} H_r {r}^{3}+2\,{r}^{5}H_r+3\,H {
a}^{4}+9\,H {N}^{2}{r}^{2}+6\,H  
{r}^{4}}{ 6 f^{7/4}H ^{4/3}{r}^{5}} },\\
\omega_{\hat r\hat z \hat x}&=&-\frac{N^2\sin\theta\cos\theta}{(N^2+r^2)^{1/4}H^{1/3}r^{1/2}(N^2\cos^2\theta+r^2)},\\
\omega_{\hat z\hat r\hat r}&=&-\frac{N^2(N^2\cos^2\theta+2r^2\cos^2\theta-r^2)}{2(N^2+r^2)^{3/4}r^{3/2}(N^2\cos^2\theta+r^2)H^{2/3}}.
\end{eqnarray}
We find the components of $\omega_{M \nu \rho}$  from equation (\ref{omeg2}). There are 59 non-zero components. We list here a few non-zero  $\omega_{M \nu \rho}$, which are given by
\begin{eqnarray}
\omega_{ t\hat y\hat t}&=&-\frac{H'_y}{6H^{3/2}},\\
\omega_{ t \hat r \hat t}&=&-\frac{f^{1/4}H'_r}{6H^{3/2}},\\
\omega_{y\hat r \hat r}&=&-\frac{H'_y}{3H},
\end{eqnarray}
\begin{eqnarray}
\omega_{ r\hat \theta \hat \theta}&=&\frac{-2N^2rH'_r-2r^3H'_r-3N^2H-6r^2H}{6r(r^2+N^2)H},\\
\omega_{ r\hat z \hat x}&=&-\frac{N^2\sin\theta\cos\theta}{(r^2+N^2)^{1/2}(r^2+N^2\cos^2\theta)},\nonumber\\
&&\\
\omega_{\theta\hat z\hat z}&=&\frac{N^2\sin\theta\cos\theta}{r^2+N^2\cos^2\theta}.\label{o125}
\end{eqnarray}
Collecting all the terms in the Killing spinor equation (\ref{K}) with $M=t$, and assuming the spinor $\epsilon$ is time-independent 
($\partial _t \epsilon=0$), we find
\be
{\Sigma}_1(\Gamma_{\hat t \hat y}+\Gamma_{\hat t \hat r\hat \theta \hat x \hat z})\epsilon+{\Sigma}_2(\Gamma_{\hat t\hat r}+\Gamma_{\hat t\hat y\hat \theta \hat x \hat z})\epsilon=0,\label{Kt5}
\ee
where
\be
{\Sigma}_1=-\frac{1}{12H^{9/6}(y,r)}\left( {\frac {
\partial }{\partial y}}H \left( y,r \right)  \right),
\ee
and
\be
{\Sigma}_2=\frac {(r^2+{a}
^{2})^{1/4}}{12 r^{3/4}H^{9/6}(y,r)   }  \left( {\frac {
\partial }{\partial r}}H \left( y,r \right)  \right).
\ee
Quite interestingly similar to M2-brane, we notice the dependence of the Killing spinor equation to the metric function and its derivative are contained in two independent functions ${\Sigma}_1$ and ${\Sigma}_2$. Multiplying equation (\ref{Kt5}) from left by $\Gamma_{\hat t}$ and from right by $\Gamma_{\hat y}$ leads to 
\be
(\Gamma_{\hat r\hat \theta\hat x \hat z\hat y}-1)\epsilon=0,\label{firstt5}
\ee
which is exactly the projection equation (\ref{first5}), since 
\be
\Gamma_{\hat r\hat \theta\hat x \hat z\hat y}=\Gamma_{\hat t\hat x_1\hat x_2 \hat x_3\hat x_4 \hat x_5}.
\ee
Similar calculations show the Killing spinor equation (\ref{K}) with $M=x_1,\cdots, x_5$ lead to the same equation (\ref{firstt5}) or (\ref{first5}). Moreover, we notice that projection equation (\ref{firstt5}) is independent of the integral equation (\ref{GENFFIN5}) for the metric function.  
The Killing spinor equation (\ref{K}) with $M=y$ has two contributions from the second term and third and fourth terms. The contribution from the second term in (\ref{K}) is given by
\be
-\frac{1}{6r^{1/2}H(y,r)}(N^2+r^2)^{1/4}\left(\frac {
\partial }{\partial r}H(y,r)\right)\Gamma_{\hat r\hat y}\epsilon.\label{Ky5}
\ee
The contribution from the third and fourth terms in (\ref{K}) is equal to 
\be
-\frac{1}{6r^{1/2}H(y,r)}(N^2+r^2)^{1/4}\left(\frac {
\partial }{\partial r}H(y,r)\right)\Gamma_{\hat y\hat r}\epsilon.\label{Ky25}
\ee
Hence the Killing spinor equation (\ref{K}) with $M=y$, is trivially satisfied. A similar calculation shows that the Killing spinor equation (\ref{K}) with $M=r$ is also trivially satisfied.
The Killing spinor equation (\ref{K}) with $M=\theta$ is
\be
\partial _{\theta}\epsilon+{\Theta}_1(\Gamma_{\hat y \hat \theta}+\Gamma_{})\epsilon+{\Theta}_2(\Gamma_{\hat r\hat \theta}+\Gamma_{})\epsilon+{\Theta}_3\Gamma_{\hat r\hat \theta}\epsilon+{\Theta}_4\Gamma_{\hat x\hat z}\epsilon=0,\label{Ka15}
\ee
where the four independent functions ${\Theta}_i,\,i=1,\cdots ,4$ are given by
\be
{\Theta}_1=-\frac{r^{1/2}(N^2+r^2)^{1/4}}{6H(y,r)} \left( {\frac {
\partial }{\partial y}}H \left( y,r \right)  \right),
\ee
\be
{\Theta}_2=-\frac{r(N^2+r^2)^{1/2}}{6H(y,r)} \left( {\frac {
\partial }{\partial r}}H \left( y,r \right)  \right),
\ee
and
\be
{\Theta}_3=-
{\frac { \left(  \left( {N}^{2}+2\,{r}^{2} \right)   \cos^2
 \theta +{r}^{2} \right) {N}^{2}}{\sqrt {r} \left( 4\,  \cos^2 \theta 
 \sqrt {{N}^{2}+{r}^{2}}\,{N}^{2}+{r}^{2} \right) }},
\ee
and
\be
\Theta_4=- {\frac { \left(  \left( {N}^{2}+2\,{r}^{2}
 \right)   \cos ^2 \theta -{r}^{2}
 \right) {N}^{2}}{\sqrt {r} \left( 4\,  \cos ^2 \theta
 \sqrt {{N}^{2}+{r}^{2}}\,{N}^{2}+{r}^{2} \right) }
}.
\ee
Using equation (\ref{firstt5}) in (\ref{Ka15}), we find
\be
\partial _{\theta}\epsilon+{\Theta}_3\Gamma_{\hat r\hat \theta}\epsilon+{\Theta}_4\Gamma_{\hat x\hat z}\epsilon=0,
\ee
which is exactly equation (\ref{M5theta}). Similar calculations show the Killing spinor equation (\ref{K}) with $M=x$ and $M=z$ leads to equations (\ref{M5x}) and (\ref{K5M5}). 
}}


\end{document}